\documentclass[11pt]{article}
\usepackage{amsmath,amssymb,amsfonts}
\usepackage[T2A]{fontenc}
\newtheorem{lemma}{Lemma}
\newtheorem{theorem}{Theorem}
\newtheorem{proposition}{Proposition}

\def\D{{\cal D}}
\def\R{{\mathbb R}}
\def\C{{\mathbb C}}
\def\Z{{\mathbb Z}}

\def\Im{{\mathrm{Im}\, }}

\def\const{{\mathrm{const}}}

\def\diag{\mathrm{diag}\,}
\def\res{{\mathrm{res}}\,}
\def\nm{{\mathrm{nm}}}
\def\Sing{{\mathrm{Sing}\,}}
\def\sgn{{\mathrm{sgn}\,}}
\def\x{{\bf{x}}}

\def\beq{\begin{equation}}
\def\eeq{\end{equation}}
\def\bet{\begin{theorem}}
\def\eet{\end{theorem}}
\def\bi{\begin{itemize}}
\def\ei{\end{itemize}}

\begin{document}
\title{Singular spectral curves in finite gap integration}
\author{Iskander A. TAIMANOV
\thanks{Institute of Mathematics, 630090 Novosibirsk, Russia;
e-mail: taimanov@math.nsc.ru}}
\date{}
\maketitle

\tableofcontents

\section*{Introduction}

This article is an extended version of the talk given by the author at the
conference ``Geometry, Dynamics, Integrable Systems
--- GDIS 2010'' (Serbia, September 7--13, 2010) dedicated to the 60th birthdays of B.A. Dubrovin,
V.V. Kozlov, I.M. Krichever, and A.I. Neishtadt.

In the talk as well as in the text we restrict ourselves
to a description of a pair of examples studied by us together with
A.E. Mironov \cite{MT1,MT2} and P.G. Grinevich \cite{GT1,GT2} and which demonstrate how
in integrable problems from differential geometry there naturally appear
singular spectral curves:

\begin{itemize}
\item
{\sl in the construction of finite gap orthogonal curvilinear coordinate sys\-tems
and solutions of the associativity equations \cite{Krichever97}
it is naturally to consider the degenerate case when the geometrical genus
of a singular spectral curve is equal to zero \cite{MT1,MT2}.}\\
In the case the construction of the Baker--Akhiezer function and of finite gap solutions
is reduced to solving linear systems, solutions are expressed in terms of elementary
functions and, in particular, one may obtain solutions, to the associativity equations,
satisfying the quasihomogeneity condition which gives a construction of infinitely many
 unknown before Frobenius manifolds.
\end{itemize}

\begin{itemize}
\item
{\sl it appears that for soliton equations with self-consistent sources
the spectral curve may be deformed and a deformation reduces to creation and annihilation of
double points \cite{GT1,GT2}.}
\end{itemize}

We recall the main notions in \S 1,
expose examples 1 and 2 in \S 2 and \S 3 respectively and briefly
note in \S 4 more interesting examples of integrable problems
in which singular spectral curves appear.

\section{Baker--Akhiezer functions on singular spectral cur\-ves}
\label{s1}

\subsection{Spectral curves}
\label{subsec1.1}

The definitions of spectral (complex) curves
\footnote{Topologically these are Riemann surfaces however, since the finite gap integration method
deals with finite genus curves, it uses the terminology from complex algebraic geometry.}
may be splitted into three types which are related to each other:

1) for one-dimensional differential operators these curves are defined via Bloch functions and
first that was done for Schr\"odinger operators by Novikov \cite{Novikov1974}.
This procedure includes an explicit construction of Bloch functions;

2) for two-dimensional differential operators the spectral curves are defined
implicitly for a fixed energy level via the ``dispersion'' relation between qua\-si\-mo\- menta.
These curves were introduced by Dubrovin, Krichever, and Novikov
\cite{DKN} who did that for the Schr\"odinger operator in a magnetic field
\beq
\label{magnetic}
L = \partial\bar{\partial} + A \bar{\partial} + U.
\eeq
In \cite{DKN} the inverse problem of reconstruction of such operators from algebro-geometrical spectral data
which include the spectral curve was also solved for operators which are
finite gap on a fixed energy level;

3) for problems which are integrated by using Baker--Akhiezer functions,
introduced by Krichever \cite{Krichever1977}, the spectral curves are
Riemann surfaces on which these functions are defined.
\footnote{This includes the case when the spectral curve
$$
P(z,w)=0, \ \ \ P \in \C[z,w],
$$
is defined algebraically via the Burchnall--Chaundy theorem which reads that
commuting ordinary differential operators $A_1$ and $A_2$ meet an algebraic relation
$P(A_1,A_2)=0$.}

We briefly expose the main notions referring for details to surveys
\cite{DMN,Krichever1977,KN,Dubrovin1981}.

1) In the finite gap integration spectral curves first appear in the initial article by Novikov
\cite{Novikov1974} as Riemann surfaces which parameterize Bloch functions of the operator
$L = -\frac{d^2}{dx^2} + u(x)$ which comes into the Lax representation
$$
\frac{\partial L}{\partial t} = [L,A]
$$
for the Korteweg--de Vries (KdV) equation. A Bloch function of an operator $L$  with a periodic potential
$u(x) = u(x+T)$ is a joint eigenfunction of $L$ and of the translation operator $\widehat{T}$: $\widehat{T}(f)(x) = f(x+T)$,
i.e., a solution of the system
\beq
\label{bloch}
L \psi = E\psi, \ \ \ \ \widehat{T}\psi = e^{ip(E)T}\psi.
\eeq
Since $L$ is of order two, for every value of $E$ it has a two-dimensional space $V_E$
formed by all solutions of the equation $L\psi = E\psi$ and this space is invariant under $\widehat{T}$.
Therewith {\it the spectral curve} $\Gamma$ appears as a two-sheeted cover
$$
\pi: \Gamma \to \C,
$$
onto which the Bloch function $\psi(x,P)$ is correctly defined as a meromorphic function in $P \in \Gamma$,
i.e. $\Gamma$ parameterizes all Bloch functions. Since $\det \widehat{T}\vert_{V_E} =1$, the eigenvalues of
$\widehat{T}\vert_{V_E}$ are equal to $\pm 1$ at branch of this covering and
$\widehat{T}$ is not diagonalized at these points. If there are finitely many such points then the operator $L$ is called
{\it finite gap},
\footnote{This terminology --- finite gap operators and Bloch functions --- came from
solid state physics. Later in English the term of ``finite zone'' was replaced by ``finite gap''.}
and the Riemann surface $\Gamma$ is completed up to a (complex) algebraic curve by a branch point at infinity
$E=\infty$.

Let us recall that the KdV equation is the initial equation of a hierarchy of equations which are represented in the form
$$
\frac{\partial L}{\partial t_k} = [L,A_k]
$$
where $A_k$ is a differential operator of order $2k+1$ and the flows corresponding to these equations do commute.

In particular, Novikov proved two fundamental results which lies in the basement of the finite gap integration method
\cite{Novikov1974}:

\begin{itemize}
\item
{\sl if $u(x)$ meets the equation of the form
$$
[L,A_N + c_1 A_{N-1} + \dots + c_{N-1}A_1] = 0
$$
(Novikov's equation), then the spectral curve is a Riemann surface of the form
$w^2 = Q(E)$ with a polynomial $Q(E)$ of degree $2N+1$
(such solutions of the KdV equation are also called {\it finite gap});

\item
the spectral curve $\Gamma$ and the quasimomentum function $p(E)$ which is correctly defined on $\Gamma$
up to $\frac{2\pi k}{T}, k \in \Z$, are first integrals of the KdV equation. In particular, for the roots of $Q(E)=0$
are first integrals of finite-gap solutions and the classical Kruskal--Miura integrals are described in terms of the branch points.}
\end{itemize}

For smooth real-valued solutions of the KdV equation the spectral curve is non-singular and real:
the polynomial $Q(E)$ has no multiple roots and all roots are real.

Analogously there are defined the spectral curves for all one-dimensional (scalar and matrix) differential and difference
operators with periodic coefficients. The construction of the spectral curve reduces to a solution of
ordinary differential equations to which the equation  $L\psi= E\psi$ (see (\ref{bloch})) reduces.

2) For two-dimensional differential operators $L$ with periodic coefficients we define
{\it Floquet functions} as solutions of the following problem:
$$
L\psi = E\psi, \ \
\psi(x+T_1,y) = e^{ip_1 T_1}\psi(x,y), \ \
\psi(x,y+T_2)
e^{ip_2 T_2} \psi(x,y),
$$
where $T_1$ and $T_2$ are the periods, and $p_1$ and $p_2$ are the quasimomenta.
{\it The multipliers} Floquet functions are defined as $e^{ip_1 T_1}$ and $e^{ip_2 T_2}$.

For hypoelliptic operators (for instance, for
$\partial\bar{\partial} + \dots$ and $\partial_y -
\partial^2_x + \dots$ where we denote by points the terms of lower order) one may show
that the energy $E$ and the quasimomenta $p_1$ and
$p_2$ meet an analytical ``dispersion'' relation
\beq
\label{curve}
F(E,p_1,p_2) = 0.
\eeq
The complex curve in the $p$-space defined by (\ref{curve}) for a fixed value of
$E$ is called {\it the spectral curve on the energy level $E$}
\cite{DKN}. The existence of such a  curve (possibly of infinite genus) can be established
either by perturbation method \cite{Krichever1989}, either by using the Keldysh theorem
\cite{Taimanov2006},

In particular, by using the Weierstrass representation, for tori in $\R^3$ (and in $\R^4$),
via solutions to the equation $\D \psi = 0$ with $\D$ a two-dimensional Dirac operator with
periodic coefficients in
\cite{Taimanov1998} there was introduced the spectral curve of a torus,
immersed into $\R^3$, as the spectral curve of the operator $\D$ on the zero energy level.
It appeared that this spectral curve reflects the geometrical properties of the surface
\cite{Taimanov2006} (see also \cite{Taimanov1997}). In \S
\ref{subsec4.2} we discuss an example related to spectral curves
of tori.

In difference with the one-dimensional situation the spectral curve
can not be constructed by solution of the direct problem. However it enters
into the inverse problem data for the Baker--Akhiezer function method.

3) Let us demonstrate a solution of the inverse problem
and a construction of solutions of nonlinear equations by using Baker--Akhiezer functions
as that was done initially for the Kadomtsev--Petviashvili (KP) equation.

By developing methods, by Dubrovin and Its--Matveev,
of constructing finite gap solutions of the KdV equations,
Krichever defined the {\it Baker--Akhiezer function}
(for the KP equation) and by using it solved the inverse problem of
constructing solutions to the KP equation from the algebro-geo\-met\-ri\-cal spectral data as follows
\cite{Krichever1976}.

Let $\Gamma$ be a smooth Riemann surface of genus $g$, $P$ be a fixed point on it
and $k^{-1}$ be a local parameter near this point such that $k(P) = \infty$, and let $D =
\gamma_1+\dots+\gamma_g$ be a generic effective divisor of degree $g$
($\gamma_1,\dots,\gamma_g \in \Gamma$). Then

\begin{itemize}
\item
{\sl there exists a unique function $\psi(x,y,t,\gamma)$,
$\gamma \in \Gamma$ which meromorphic in $\gamma$ on $\Gamma \setminus
P$, has poles only at points from $D$ (more precise, divisor of poles
$\leq D$) and has an asymptotic
$$
\psi = e^{kx+k^2y+k^3t}\left(1+ \sum_{m>0}
\frac{\xi_m(x,y,t)}{k^m}\right) \ \ \mbox{as $\gamma \to P$};
$$
The function $u(x,y,t)= -2\partial_x \xi_1$
satisfies the KP equation
$\frac{3}{4}\,u_{yy} = \frac{\partial}{\partial x}\left[u_t -
\frac{1}{4}\left(6uu_x + u_{xxx}\right)\right]$,
for all $\gamma \in \Gamma \setminus P$ we have
\beq
\label{spectral}
\frac{\partial \psi}{\partial y} = L\psi, \ \ \mbox{где $L = \frac{\partial^2}{\partial x^2}+u$},
\eeq
and there is an explicit formula for
$u(x,y,t)$ in terms of the theta function of $\Gamma$:
\beq
\label{its}
u(x,y,t) = 2 \frac{\partial^2}{\partial x^2} \log \theta (Ux+Vy+Wt+z_0)+c.
\eeq
}
\end{itemize}

The set $(\Gamma,P,k^{-1},D)$ defines Krichever's ``spectral data'' for the inverse problem for solutions to the KP equation and for the operator
$\frac{\partial}{\partial y}-L$, for which, by (\ref{spectral}),
$\psi$ is a Floquet function on the zero energy level in the case when the potential $u$ is periodic.

If on $\Gamma$ there exists a meromorphic function with a unique pole which is of the second order and at the point $P$, then there exists a two-sheeted covering $\Gamma \to \C P^1$  which is ramified at $P$, the function
$u$ does not depend on $y$, $\psi$ reduces to the Bloch function of the Schr\"odinger operator $L$,
the formula (\ref{its}) reduces to the Its--Matveev formula for finite gap solutions of the KdV equation, and
the set $(\Gamma,P,k^{-1},D)$ becomes the inverse problem data for a finite gap Schr\"odinger equation.
These results for the KdV equation were obtained before in the articles by Dubrovin, Novikov, Its, and Matveev
(see the survey \cite{DMN}).

A Riemann surface $\Gamma$ is called {\it the spectral curve},
and solutions obtained by this method are called {\it finite gap}. Therewith
quasiperiodic solutions are not excluded (it is easy to notice from
(\ref{its}) that such solutions may be obtained by using this method).
This terminology is used for all problems which are solved by using Baker--Akhiezer functions
(by using {\it the finite gap integration method}).

Real-valued periodic solutions of the KdV, KP, and sine-Gordon equations are constructed from smooth Riemann surfaces. Theta functional formulas are rather complicated because parameters coming into them are related by
complex transcendent equations.

At the same time for some equations to which finite gap integration is applied interesting solutions are
constructed from singular spectral curves including curves with geometrical genus zero.
In the latter case a solution is expressed in terms of elementary functions and admits a simple qualitative
investigation  (see \S\S 2 and 3).

\subsection{Algebraic curves with singularities}
\label{subsec1.2}

Let us recall the main notions related to complex algebraic curves with singularities
(singular curves; see details in
\cite[Chapter 4]{Serre}).

Let $\Gamma$ be a complex algebraic curve with singularities.

Then there exists a morphism of a smooth algebraic curve
$\Gamma_\nm$:
$$
\pi: \Gamma_\nm \to \Gamma,
$$
such that

1) in $\Gamma_\nm$ there is a finite set of points $S$, splitted into subclasses and $\pi$
maps $S$ exactly into the set of singular points
$\Sing = \Sing \Gamma$ of $\Gamma$ and moreover the preimage of every point from
$\Sing$ consists in a certain subclass of $S$;

2) the mapping $\pi: \Gamma_\nm \setminus S \to \Gamma \setminus
\Sing$ is a smooth one-to-one projection;

3) every regular mapping $F: X \to \Gamma$ of a nonsingular algebraic variety
$X$ with everywhere dense image $F(X)
\subset \Gamma$ descends through $\Gamma_\nm$, i.e $F = \pi G$
for some regular mapping $G: X \to \Gamma_\nm$.

A mapping $\pi$ satisfying these conditions is called
{\it the normalization} of $\Gamma$ and is uniquely defined by the conditions.
Sometimes the curve $\Gamma_\nm$ itself is called the normalization.

The genus of $\Gamma_\nm$ is called {\it the geometrical genus} of
$\Gamma$ and is denoted by $p_g(\Gamma)$.

Into the Riemann--Roch formula there enters {\it the arithmetic genus}
$p_a(\Gamma)$ which is a sum of the geometric genus and a positive contribution of
singularities (the points from $\Sing$). For a smooth curve we have $p_a = p_g$.

A meromorphic $1$-form $\omega$ on $\Gamma_\nm$ defines {\it
a regular differential} on $\Gamma$ if for every point $P \in
\Sing$ we have
$$
\sum_{\pi^{-1}(P)} \mbox{Res}\ (f\omega) = 0
$$
for all meromorphic functions $f$, on $\Gamma_\nm$, which descends to
functions on $\Gamma$, i.e have the same value at points from every divisor
$D_i$ and have no poles at $\pi^{-1}(P)$.
Regular differentials may have poles at the preimages of singular points. The dimension
of the space of regular differentials is equal to
$p_a(\Gamma)$.

Let us take on an irreducible algebraic curve $\Gamma_\nm$
$s$ families $D_1,\dots,D_s$, consisting in $r_1,\dots,r_s$ points all of which
are different and let us construct $\Gamma$ by identifying all points from the same family.
Then
$$
p_a(\Gamma) = p_g(\Gamma) + \sum_{i=1}^s (r_i-1).
$$

Let us formulate the Riemann--Roch theorem for algebraic curves with singularities:

\begin{itemize}
\item
{\sl Let $L(D)$ be a space of meromorphic functions on $\Gamma$ with
from $D = \sum n_P P$ of order $\leq n_P$ and
$\Omega(D)$ be a space of regular differentials, on $\Gamma$,
which has at every point $P \in \Sing$ a zero of order not less than $n_P$. Then
$$
\dim L(D) - \dim \Omega(D) = \deg D + 1 - p_a(\Gamma).
$$
For generic divisor $D$ with $\deg D \geq p_a$ we have
$\dim \Omega(D)=0$ and
$$
\dim L(D) = \deg D + 1 - p_a(\Gamma).
$$
}
\end{itemize}

The standard scheme of proving the uniqueness of the Baker--Akhiezer function is based on the
Riemann--Roch theorem ad the genus $g$ of a smooth spectral curve $\Gamma$ comes into all
reasonings as the arithmetic genus. For the case of singular spectral curves it is enough
to replace $g$ by $p_a$ in all arguments and definitions.

Let us expose the simplest examples of one-dimensional finite gap
Schr\"o\-din\-ger operators with singular spectral curves.

{\sc Example 1 (a curve with a double point).}
Let $\Gamma_\nm = \C P^1 = \C \cup \{\infty\}$ with a parameter
$k$. We construct a curve $\Gamma$ by identifying the points $k=\pm \lambda$ on $\Gamma_\nm$.
We have $p_a(\Gamma)=1$ and therefore put $D = \{k=0\}$. From the spectral data
$(\Gamma,\infty,k,D)$ there are constructed the Schr\"odinger operator
\footnote{Its from differs from (\ref{spectral}) by a sign at
$d^2/dx^2$ which is reflected in the change of the asymptotic of $\psi$: $\psi \approx e^{ikx}$
as $k \to \infty$.}
$$
L = -\frac{d^2}{dx^2} + \frac{2\lambda^2}{\sin^2\,(\lambda x)}
$$
and the Baker--Akhiezer function
$$
\psi(k,x) = \left(1 + \frac{i\lambda}{k}\frac{\cos(kx)}{\sin(kx)}\right)e^{ikx}.
$$

{\sc Example 2 (a cuspidal curve).}  Let $\Gamma$ be the same as in the previous example.
The curve $\Gamma$ which homeomorphic to $\Gamma_\nm$ but has a cuspidal sin\-gu\-la\-rity at
$k=0$ is obtained by restricting of the class of locally holomorphic functions on
$\Gamma_\nm$: we assume that a function $f$ which is holomorphic near $k=0$
on $\Gamma_\nm$ is holomorphic in a neighborhood of $k=0$ on $\Gamma$ if and only if
\beq
\label{cusp}
f^\prime\vert_{k=0} = 0.
\eeq
We have $p_a(\Gamma)=1$.
The spectral data $(\Gamma,\infty,k,D = \{k=0\})$ defines the potential
$$
u(x) = \frac{2}{x^2},
$$
and the function
$$
\psi_1(k,x) = \left(-\frac{\partial}{\partial x} + \frac{1}{x}\right) e^{ikx} = \left(-ik+\frac{1}{x}\right)e^{ikx}
$$
parameterizes all Bloch function and satisfies (\ref{cusp}). After diving it by $-ik$
we obtain the Baker--Akhiezer function normalized by the asymptotic:
$$
\psi(k,x) = \left(1 +\frac{i}{kx}\right)e^{ikx} \ \ \ \ \mbox{as $k \to \infty$}.
$$
This potential is obtained from the potentials described in Example 1 in the limit
as $\lambda \to 0$.

{\sc Example 3.} The potential from Example 2 enters a series of rational soliton potentials:
\beq
\label{hcusp}
u_l(x) = \frac{l(l+1)}{x^2}.
\eeq
The Baker--Akhiezer function for an operator with the potential
(\ref{hcusp}) is defined on the curve $\Gamma$ which is obtained from
$\Gamma_\nm = \C P^1$ by assuming the condition which is analogous to
(\ref{cusp}):
$$
f^\prime = f^{\prime\prime} = \dots = f^{(l)} = 0 \ \ \ \mbox{при $k=0$}.
$$
We have $p_a(\Gamma) = l$, $D = lQ$, where $Q = \{k=0\}$, and
another spectral data are the same as in Example 2.
The Baker--Akhiezer function
has the form
$$
\psi_l(k,x) = \frac{1}{(-ik)^l}\left(-\frac{\partial}{\partial x} + \frac{l}{x}\right)\dots
\left(-\frac{\partial}{\partial x} + \frac{1}{x}\right) e^{ikx}.
$$
The Riemann surface $\Gamma$ is defined by the equation $y^2 = x^{2l+1}$ and the normalization has the form
$$
\C \to \Gamma \setminus\{\infty\}, \ \ \ t \to (x = t^2, y = t^{2n+1}).
$$

Usually the Baker--Akhiezer function is constructed as a function
$\psi$ on $\Gamma_\nm$ and multiple points and cuspidal singularities are
defined by additional conditions
$$
\psi(Q_1)=\psi(Q_2), \ \ \ \psi^\prime (Q) = \dots = \psi^{(l)}(Q).
$$
At the same time in soliton theory there were considered more general
constraints of the form
\beq
\label{sing}
\sum_{j=1}^M \sum_{i=1}^{m_j}
a_{ij} \psi^{(j)}(k,x,\dots)\vert_{k=Q_j} = 0, \ \ \ Q_1,\dots,Q_M
\in \Gamma_\nm, \eeq
or even systems of such conditions (see, for instance, \cite{DKMM}).
In this case $\psi$ is a section of a certain bundle over
the curve $\Gamma$ which we may determine as follows.
It is evident that generically a constraint of the form
(\ref{sing}) posed on rational functions on
$\Gamma_\nm$ does not distinguish any subfield:
a product of two functions meeting such a condition does not satisfy it.
At the same time sections of the bundle which we are looking for
form a module over the field of rational functions on
$\Gamma$. Let $f$ be such a function and $\psi$ be a section of the bundle
defined by (\ref{sing}), then the minimal conditions to which $f$ has to satisfy
for $f\psi$ to satisfy (\ref{sing}) for all sections are as follows:
$$
f(Q_1) = \dots = f(Q_M), \ f^\prime(Q_j) = \dots = f^{(m_j)}(Q_j) = 0, \ \ j=1,\dots,M,
$$
i.e. $\Gamma$ is obtained from $\Gamma_\nm$ by gluing points $Q_1,\dots,Q_M$
and therewith every such a point with
$m_j \geq 1$ is singular (see Example 1).

However some of these functions (apparently all of them) may be obtained from functions on smooth spectral curves
by degenerations of curves. Let expose the following

{\sc Example 4.}\, (A.E. Mironov) \, Let $\Gamma_1$ be a connected algebraic curve of genus
$g$ on which there is constructed the Baker--Akhiezer function $\psi$. For simplicity we assume that
$\psi$ is constructed from the spectral data for the KP equation however it satisfies an
additional condition
$$
\psi(Q_1) = \alpha \psi(Q_2),
$$
where $\alpha \neq 1$. By the Riemann--Roch theorem, the fixing of poles divisor
of $\psi$ (it consists in $g+1$ generic points
$\gamma_1,\dots,\gamma_{g+1}$ taken together with their multiplicities)
together with the asymptotic at the essentially singular point determines
$\psi$ uniquely. Let us consider a rational curve $\Gamma_2
= \C P^1$ with marked points $a$ and $b$, which differ from $\infty$,
and let us identify $Q_1$ with $a$ and $Q_2$ with $b$ to obtain a reducible curve
$\Gamma = \Gamma_1 \cup \Gamma_2 / \{Q_1
\sim a, Q_2 \sim b\}$ with the arithmetic genus $p_a(\Gamma)=g+1$.
Let us construct on $\Gamma$ the Baker--Akhiezer function
$\widetilde{\psi}$ with $g+2$ poles at $\gamma_1,\dots,\gamma_{g+1} \in \Gamma_1$ and
at the point $q \in \Gamma_2$ such that $(b-q)/(a-q) = \alpha$. The number
of poles is greater by one than the arithmetic genus and we pose
an additional normalization condition
$\psi(\infty) = 0$ where $\infty \in \Gamma_2$.
Then the restriction of $\psi$ onto $\Gamma_2$ is the function
$\varphi = \frac{c}{z-q}, c = \mathrm{const} \neq 0$.
By the choice of $q$, we have $\varphi(a) = \alpha \varphi(b)$ and
therefore the restriction of $\widetilde{\psi}$ onto $\Gamma_1$
is the desired function $\psi$.
The Riemann surface $\Gamma$ is obtained by a degeneration of smooth surfaces
via pinching a pair of contours into the points
$Q_1 \sim a$ and $Q_2 \sim b$ and $\widetilde{\psi}$ is a degeneration of
the Baker--Akhiezer functions on these surfaces. Since the inverse problem
is solved from the asymptotics of $\psi$ at the essential singularities
and there are no such singularities on $\Gamma_2$, then the restriction of
$\widetilde{\psi}$ onto $\Gamma_2$ does not play any role. At the same time
the real spectral curve on which $\widetilde{\psi}$
is defined as the Baker--Akhiezer function is $\Gamma$ and not $\Gamma_1$.

Apparently one may obtain by this procedure the potential Schr\"odinger
operators from \cite{Taimanov2003},
\footnote{These examples were constructed at the beginning of 1990s and at the same time
Malanyuk argued that their spectral curves are obtained via degenerations from smooth curves.}
which are constructed from singular spectral curves with odd arithmetic genus
(for smooth spectral curves in this problem the genus is always even
\cite{Krichever1989}).

In \cite{Malanyuk} Malanyuk constructed finite gap solutions of the KPII equation
for which the spectral curve is reducible and the essential singularity
of the Baker--Akhiezer function lies in the rational component.
These solutions are nonlinear superpositions of soliton waves.

It is easy to notice that the construction from Example 4 is generalized
for all conditions of the form
$\alpha_1\psi(Q_1) + \dots + \alpha_M\psi(Q_M)=0$ with generic coefficients $\alpha_1,\dots,\alpha_M$.
Therewith $\Gamma$ is obtained from $\Gamma_1$ by adding a rational curve $\C P^1$ which intersects
$\Gamma_1$ at $Q_1,\dots,Q_M$.

\section{Orthogonal curvilinear coordinate systems and Frobenius manifolds}
\label{s2}

\subsection{Orthogonal curvilinear coordinates and integrable systems of hydrodynamical type}
\label{subsec2.1}

Into theory of integrable systems orthogonal curvilinear
coordinates came from two sides: they do naturally appeared in the integrability
problem for one-dimensional systems of Dubrovin--Novikov hydrodynamical type
\cite{DN83,N85,DN89} and they also do appear as a result of a differential
reduction as a particular case of metrics with diagonal curvature
to whose explicit construction the inverse problem method was
applied by Zakharov \cite{Z}.

{\it Orthogonal curvilinear coordinate system} in a Riemannian
manifold is a coordinate system
$(u^1,\dots,u^N)$ such that in these coordinates the metric tensor
takes the diagonal form
\begin{equation}
\label{diagonal}
ds^2 = \sum_{i=1}^N H_i^2(u) \left(du^i\right)^2.
\end{equation}
We note that for $N \geq 4$ not every Riemannian metric admits locally orthogonal curvilinear coordinates.
In fact, the existence of such coordinates is a very strong condition on a metric.

The coefficients $H_i$ are called {\it the Lame coefficients}, and the expressions
\begin{equation}
\label{rotation}
\beta_{ij} = \frac{1}{H_j}\frac{\partial H_i}{\partial u^j}
\end{equation}
define {\it the rotation coefficients}.

A metric is called of {\it Egorov} type if it admits locally
orthogonal curvilinear coordinates with symmetric rotation coefficients:
$$
\beta_{ij} = \beta_{ji}, \ \ \ \ 1 \leq i,j \leq N.
$$
In this case the metric is of {\it potential} type:
$$
g_{ii} = H_i^2 = \frac{\partial V}{\partial u^i}, \ \ \ i=1,\dots,N,
$$
for a certain potential $V(u)$.

In theory of hydrodynamical systems the case of Riemannian metrics is
not distinguished and there is considered the general case of pseudoriemannian metrics for which
in terms of orthogonal coordinates the metric
tensor takes the form
$$
ds^2 = \sum_{i=1}^n \varepsilon_i H_i^2(u) \left(du^i\right)^2, \ \ \varepsilon_i = \pm 1,
$$
the rotation coefficients are defined similar to the Riemannian case, and the Egorov condition
takes the form
$$
\beta_{ij} = \varepsilon_i \varepsilon_j \beta_{ji}.
$$

A one-dimensional system of hydrodynamical type is an evolution system of the form
$$
u^i_t = \sum_{j=1}^N v^i_j(u) u^j_x, \ \ \ \ i=1,\dots,N,
$$
where $u^1(x,t),\dots,u^N(x,t)$ are functions of one-dimensional spatial variable
$x$ and a temporal variable $t$. Such a system is Hamiltonian if it admits a representation of the type
$$
u^i_t = \{u^i(x), \widehat{H}\},
$$
whereе
$$
\widehat{H}(u) = \int H(u)dx
$$
and the Poisson brackets (of hydrodynamical type, or called also now
{\it Pois\-son--Dubrovin--Novikov brackets}) have the form
\begin{equation}
\label{poisson}
\{u^i(x),u^j(y)\} = g^{ij}(u(x))\delta^\prime(x-y) -
g^{is}\Gamma^j_{sk}u^k_x \delta(x-y), \ \ \ g^{ij}=g^{ji}.
\end{equation}
These notions were introduced by Dubrovin and Novikov
\cite{DN83,DN89} who proved that

1) the expression (\ref{poisson}) with a nondegenerate pseudoriemannian metric
$g^{ij}$ (with upper indices) in the $N$-dimensional $u$-space
defines Poisson brackets if and only if the metric is flat (has zero curvature)
and $\Gamma^i_{jk}$ is the corresponding Levi-Civita connection;

2) Hamiltonian systems of hydrodynamical type do appear by averaging
such integrable systems as the KdV and
sine-Gordon equations, and the nonlinear Schr\"odinger equation (NS).

Novikov stated the conjecture that Hamiltonian systems of
hydrodynamical type with a diagonal matrix
$\left( v^i_j \right)$ are integrable. For such systems
the coordinates $u^1,\dots,u^N$,
are called {\it Riemann invariants} and the matrix $\left( g^{ij} \right)$
is also diagonal the $u$-space with orthogonal curvilinear coordinates
$u^1$,\dots,$u^N$. This conjecture was proved by Tsarev who also
introduced the integration procedure --- the generalized godograph method
\cite{Tsarev}.

In theory of integrable systems an interest to Egorov metrics is due to Dubrovin
who showed that if a flat metric
$g^{ij}$, written in terms of Riemannian invariants, is of Egorov type then the system
is superintegrable  \cite{Dubrovin90}.
\footnote{For the modern state of theory of such systems we refer to \cite{Pavlov}
and references therein.}

In \cite{MF} the following nonlocal generalization of Poisson--Dubrovin--No\-vi\-kov brackets
was proposed:
$$
\{u^i(x),u^j(y)\} = g^{ij}(u(x))\delta^\prime(x-y) -
g^{is}\Gamma^j_{sk}u^k_x \delta(x-y) + \frac{c}{2} \sgn (x-y) u^i_x u^j_y,
$$
where $c = \mathrm{const}$, and it was proved that this expression
defines Poisson brackets if and only if
$\Gamma^i_{jk}$ is the Levi-Civita connection for the metric
$g^{ij}$ with constant sectional curvature $c$ and that the generalized
godograph method is applicable to Hamiltonian systems with such Poisson brackets.
Therewith Riemann invariants define orthogonal curvilinear coordinates in the space
of constant curvature $c$.

\subsection{The Lame equations}
\label{subsec2.2}

In the 19th century the theory of orthogonal curvilinear coordinates
was actively studied by leading geometers (Dupin, Gauss, Lame, Bianchi, Darboux).
The classification problem for these coordinates was basically solved up to the beginning
of the 20th century and the state of the theory for this moment was
summarized in Darboux's book \cite{Darboux}.

All such coordinate systems are constructed from solutions of the Lame equations as follows.

Let us split the nontrivial components of the Riemann curvature tensor
$R_{ijkl}$ into three groups:

1) all indices $i,j,k,l$ are pair-wise different;

2) $R_{ijik}$ with pair-wise different $i,j,k$;

3) $R_{ijij}$ with $i \neq j$.

We recall that the Riemann tensor is skew-symmetric in the first and in the second pairs of indices:
$R_{ijkl} = - R_{jikl} = -R_{ijlk}$ and $R_{ijkl} = R_{klij}$
for all $i,j,k,l$.

For diagonalized metrics (\ref{diagonal})
all components of the first type vanish and the vanishing of the components of the second type
is equivalent to the system
\beq
\label{1}
\frac{\partial^2 H_i} {\partial u^j \partial u^k} =
\frac{1}{H_j}\frac{\partial H_j}{\partial u^k}\frac{\partial
H_i}{\partial u^j} + \frac{1}{H_k}\frac{\partial H_k}{\partial u^j}
\frac{\partial H_i}{\partial u^k}, \ \ \ i\neq j \neq k,
\eeq
and the equations
\beq
\label{2}
\frac{\partial}{\partial u^j} \left(\frac{1}{H_j}
\frac{\partial H_i}{\partial u^j}\right) + \frac{\partial}{\partial
u^i} \left(\frac{1}{H_i} \frac{\partial H_j}{\partial u^i}\right) +
\sum_{k\ne i \ne j}\frac{1}{H_k^2}\frac{\partial H_i}{\partial u^k}
\frac{\partial H_j}{\partial u^k} = 0, \ \ \ i \neq j,
\eeq
are equivalent to the vanishing of the components of the third type:
$R_{ijij} = 0$.

The systems (\ref{1}) and (\ref{2}) consist in
$\frac{N(N-1)(N-2)}{2}$ and $\frac{N(N-1)}{2}$ equations, respectively.
From counting the numbers of equations and of variables it is clear that the system
(\ref{1})--(\ref{2}) on the Lame coefficients is strongly overdetermined.

The order of the system (\ref{1})--(\ref{2}) is minimized by introducing the rotation coefficients.
In this case the equations
(\ref{1}) take the form
\beq
\label{4}
\frac{\partial \beta_{ij}}{\partial u^k} = \beta_{ik}\beta_{kj}, \ \ \
i \neq j \neq k,
\eeq
and the equation (\ref{2}) are written as
\beq
\label{5}
\frac{\partial \beta_{ij}}{\partial u^i} +
\frac{\partial \beta_{ji}}{\partial u^j} + \sum_{k\ne i,j}
 \beta_{ki}\beta_{kj}=0, \ \ \ i \neq j.
\eeq

The systems (\ref{4}) and (\ref{5}) form the system of {\it the Lame equations}
and the equations (\ref{4}) are just the compatibility condition for (\ref{rotation}).
A general solution of these equations depends on $\frac{N(N-1)}{2}$ arbitrary functions of two variables.

Given a solution $\beta_{ij}$ to the Lame equations, the Lame coefficients
are found from (\ref{rotation}) as a solution to the Cauchy problem
$$
H_i(0,\dots,0,u^i,0,\dots,0)=h_i(u^i).
$$
Therewith a solution depends on the initial data
which are $N$ functions $h_i$ of one variable.

The determination of Euclidean coordinates $x^1,$
$\dots,x^N$ as functions of $u^1,\dots,u^N$ (the immersion problem) is reduced to
solving an overdetermined system of linear equations
\beq \label{6}
\frac{\partial^2 x^k}{\partial u^i \partial u^j} =
\sum_{l=1}^N\Gamma_{ij}^l\frac{\partial x^k}{\partial u^l},
\eeq
where the Christoffel symbols have the form
$$
 \Gamma_{ij}^k=0,\ i\ne j\ne k; \ \
 \Gamma_{kj}^k=\frac{1}{H_k}\frac{\partial H_k}{\partial u^j}; \ \
 \Gamma_{ii}^k=-\frac{H_i}{(H_k)^2}\frac{\partial H_i}{\partial u^k},\ i \neq k.
$$
By (\ref {1}) and (\ref{2}), the system (\ref{6}) is compatible and determines an orthogonal
curvilinear coordinate system up to motions of $\R^N$.

\subsection{Egorov metrics with zero curvature as an integrable system: Dubrovin's theorem}
\label{subsec2.3}

In \cite{Dubrovin90} Dubrovin did show that the problem of constructing flat Egorov metrics is integrated by
methods of soliton theory. We have

\begin{theorem}[\cite{Dubrovin90}]
1) A Egorov (Riemannian or pseudoriemannian) metric is flat if and only if
its rotation coefficients satisfy the equations
\begin{equation}
\label{egorov1}
\frac{\partial \beta_{ij}}{\partial u^k} = \beta_{ik}\beta_{kj}, \ \mbox{where $i,j,k$ are pair-wise different},
\end{equation}
\begin{equation}
\label{egorov2}
\sum_{k=1}^N \frac{\partial \beta_{ij}}{\partial u^k} = 0, \ \ \ i \neq j.
\end{equation}

2) The system consisting of the equations (\ref{egorov1}) and (\ref{egorov2}),
is the compatibility condition for the linear system
$$
\frac{\partial \psi_i}{\partial u^j} = \beta_{ij}\psi_j, \ \ \ i \neq j,
$$
$$
\sum_{k=1}^N \frac{\partial \beta_{ij}}{\partial u^i} = \lambda \psi_i,
\ \ \ i=1,\dots,N,
$$
where $\lambda$ is a spectral parameter.

3) For a flat Egorov metric a restriction of the rotation coefficients
$\beta_{ij}(u)$ onto every plane $u^i = a^i x + c^i t$
satisfies the $N$ waves equations
$$
[A, B_t] - [C, B_x] = [[A,B],[C,B]],
$$
где
$$
A = \diag (a^1,\dots,a^N), \ C = \diag (c^1,\dots,c^N), \
B = (\beta_{ij}),
$$
with an additional reduction
$$
\Im B =0, \ \ B^\top = JBJ, \ \ J = \diag (\varepsilon_1,\dots,
\varepsilon_N)
$$
(for Riemannian metrics all $\varepsilon_i = 1$).
\end{theorem}

This theorem is proved by straightforward computations.

\subsection{Zakharov's method of constructing metrics with diagonal curvature and
 orthogonal curvilinear coordinates}
\label{subsec2.4}

Zakharov applied the inverse scattering method to constructing a wide class
of metrics with diagonal curvature \cite{Z}. Let us expose it.

The Riemann curvature tensor is interpreted as the curvature operator on the space of tangent bivectors.
Let $M^N$ be a Riemannian (or pseudoriemannian) manifold,
$\Lambda^2 TM^N$ be a linear bundle over $M^N$ for which a general fiber over a point is the space
of bivectors at this point.
The metric on $M^N$ defines a standard metric on fibers:
if $e_1,\dots,e_N$ is an orthonormal basis for $T_x M^N$, the tangent space at
в точке $x$, then $\{e_i \wedge e_j, i < j\}$ is an orthonormal basis in $\Lambda^2 T_x M^N$.
Then the Riemann curvature tensor $R_{ijkl}$ defines the curvature operator $R$
by the formula
$$
\langle R \xi,\eta \rangle = R_{AB}\xi^A\eta^B, \ \
R_{AB} = R_{ijkl}, \ \ A = [ij], B = [kl],
$$
where  $\xi = \sum_A e_A \xi^A$ and
$\eta = \sum_B e_B \eta^B$  is a decomposition of bivectors in the basis
$e_A = e_i \wedge e_j, i<j$.
Let us remark that the well-known in the relativity theory Petrov's classification of
four-dimensional solutions to the Einstein equations is bases
on the classification of algebraic types of the curvature tensor.
\footnote{Now by Petrov's classification it is mean the analogous classification
based on the algebraic types of the Weyl operator which is analogously constructed
from the Weyl tensor (the tensor of conformal curvature).}

In \cite{Z} a manifold with diagonal curvature is defined as a manifold such that near every its
point the metric is diagonalized, i.e reduces to the form (\ref{diagonal}), and in thse coordinates
 the quadratic form $R_{AB}$ (or the curvature operator $R$) is also diagonal.

The equations (\ref{egorov1}) exactly describe the rtoation coefficients
of metrics with diagonal curvature and are written as follows:
\beq
\label{7}
 \sum_{i,j,k}\varepsilon_{ijk}\left(I_j \frac{\partial B}{\partial u^i} I_k -
 I_i B I_j B I_k \right)=0,
\eeq
where $B(u) = (\beta_{ij})$ is an $(N\times N)$-matrix function,
formed by the rotation coefficients, and $I_j = \diag
(0,\dots,0,1,0,\dots,0)$ (the unity is at the $j$-th entry).

Let us consider an auxiliary function
$\widetilde{B}=\widetilde{B}(u,s)$, where $s = u^{N+1}$
is an additional variable, which satisfies the equations
\beq
\label{8}
\frac{\partial}{\partial u^j}[I_i,\widetilde{B}] -
\frac{\partial}{\partial u^i}[I_j,\widetilde{B}] +
I_i \frac{\partial \widetilde{B}}{\partial s} I_j -
 I_j\frac{\partial \widetilde{B}}{\partial s} I_i -
 [[I_i,\widetilde{B}],[I_j,\widetilde{B}]]=0,
\eeq
where $i,j=1,\dots,N+1$ and $I_{N+1}$ is the unit matrix. If
$\widetilde{B}$ satisfies (\ref{8}), then for any fixed value of $s$
the matrix-valued function $\widetilde{B}$ satisfies the $N$ waves equation (\ref{7}).

The system (\ref{8})
admits the Lax representation
$$
[L_i,L_j]=0, \ \ \
L_j=\frac{\partial}{\partial u^j} + I_j \frac{\partial}{\partial s} + [I_j,\widetilde{B}].
$$
Let us apply to it the dressing method by considering an integral equation of the
Marchenko type
\beq
\label{9}
K(s,s^\prime,u)=F(s,s^\prime,u)+\int_s^{\infty}K(s,q,u)F(q,s^\prime,u)\,dq,
\eeq
where $F(s,s^\prime,u)$ is a matrix-valued function.

\begin{theorem}[\cite{Z}]
\label{zakharov-theorem}
1) If $F(s,s^\prime,u)$ meets the following conditions:

а) the holds the equation
\beq
\label{10}
\frac{\partial F}{\partial u^i} + I_i \frac{\partial F}{\partial s} F +
\frac{\partial F}{\partial s^\prime} F I_i = 0,
\eeq

б) the equation (\ref{9}) has a unique solution,

\noindent
then the function
\beq
\label{11}
\widetilde{B}(s,u) = K(s,s,u)
\eeq
satisfies (\ref{8}) and therefore for any fixed value of $s$ the function $B(u) = \widetilde{B}(s,u)$
satisfies (\ref{7}) and defines the rotation coefficients of a metric with diagonal curvature.
\end{theorem}

The class of metrics with diagonal curvature is rather wide and, for instance, as it was noted by Zakharov
it includes many solutions of the Einstein equations including the Schwarzschild metric.
The problem posed by him on finding a reduction, of the inverse problem data, corresponding to
Ricci-flat metrics stays an important open problem. By using a new and original trick, i.e.
{\it a differential reduction}, he distinguishes on the language of inverse problem data, i.e.
in terms of the function $F$, the class of flat metrics with diagonal curvature.

\begin{theorem}[\cite{Z}]
Let us assume the conditions of Theorem \ref{zakharov-theorem} and assume that
$F$ satisfies the relation
\beq
\label{12}
\frac{\partial F_{ij}}{\partial s^\prime}(s,s^\prime,u) +
\frac{\partial F_{ji}}{\partial s}(s^\prime,s,u)=0.
\eeq
Then $\widetilde{B}$
satisfies the equations (\ref{5}) and the rotation coefficients constructed from $\widetilde{B}$
correspond to flat metrics of the form (\ref{diagonal}), i.e. to orthogonal curvilinear coordinate systems.
\end{theorem}

The system consisting of (\ref{10}) and (\ref{12}) has the following solution \cite{Z}:

Let $\Phi_{ij}(x,y), i<j$, be $\frac{N(N-1)}{2}$ arbitrary functions of two variables and
let $\Phi_{ii}(x,y)$ be $N$ arbitrary skew-symmetric functions:
$$
\Phi_{ii}(x,y)=-\Phi_{ii}(y,x).
$$
Let us put
$$
F_{ij}=\frac{\partial \Phi_{ij}(s-u^i,s^\prime-u^j)}{\partial s},\
F_{ji}=\frac{\partial \Phi_{ij}(s^\prime-u^i,s-u^j)}{\partial s}, \
i \neq j,
$$
$$
F_{ii}=\frac{\partial \Phi_{ii}(s-u^i,s^\prime-u^i)}{\partial s}.
$$
Then $F=(F_{ij})$ satisfies (\ref{10}) and (\ref{12}), and {\sl for any fixed value of $s$
a solution $K$  of the equation (\ref{9}) with such a matrix $F$ defines the rotation coefficients
of an orthogonal coordinate system: $\beta_{ij}(u) =
K_{ij}(s,s,u)$.} Notice that we have $\frac{N(N+1)}{2}$ functional parameters $\Phi_{ij}, i \leq j$, and
a general solution depends on $\frac{N(N-1)}{2}$ functional parameters which implies that this method gives
equivalent classes of dressings.

A general theory of reductions, including differential reductions, which may be helpful for distinguishing other classes of metrics is discussed in \cite{ZM}. In particular, by a variation of the reduction
(\ref{12}) it is possible to distinguish metrics with diagonal curvature and constant sectional curvature
$K \neq 0$, i.e. orthogonal curvilinear coordinates in the spaces of constant curvature $K$ (Zakharov).
In this case the system (\ref{5}) is replaced by the equations
$$
\frac{\partial \beta_{ij}}{\partial u^i} +
\frac{\partial \beta_{ji}}{\partial u^j} + \sum_{k\ne i,j}
\beta_{ki}\beta_{kj} = -KH_iH_j, \ \ \ i \neq j.
$$

\subsection{Krichever's construction of finite gap orthogonal curvilinear coordinates}
\label{subsec2.5}

In \cite{Krichever97} Krichever proposed the finite gap version of
a construction of orthogonal curvilinear coordinates in the Euclidean spaces, which
in difference with \cite{Z} does not split into two parts: a construction  of
metrics with diagonal curvature and then distinguishing flat metrics among them
but straightforwardly gives such coordinates.

Let $\Gamma$ be a smooth complex algebraic curve, i.e. a compact smooth
Riemann surface.

Let us choose three effective divisors on $\Gamma$:
$$
 P=P_1+\dots+P_N, \ \
D = \gamma_1+\dots+\gamma_{g+l-1}, \ \
 R=R_1+\dots+R_l,
$$
where $g$ -is the genus of $\Gamma$, $P_i,\gamma_j,R_k\in\Gamma$.
Let us take a local parameter $k_i^{-1}$ near every point $P_i$, $i=1,\dots,N$,
such that this parameter vanishes at the corresponding point.

The Baker--Akhiezer function which corresponds to this spectral data
$S = \{\Gamma,P,D,R\}$ is a function $\psi(u^1,\dots,u^N,z),\ z\in\Gamma$ such that

1) $\psi\exp(- u^i k_i)$ is analytical near $P_i$,
$i=1,\dots,N$;

2) $\psi$ is meromorphic on $\Gamma\backslash\{\cup P_i\}$ with poles in
$\gamma_j$, $j=1,\dots,g+l-1$;

3) $\psi(u,R_k)=1$, $k =1,\dots,l$.

For generic divisor $D$ such a function exists, unique and is expressed
in terms of the theta-function of $\Gamma$.

If  $\Gamma$ is not connected it is assumed that the restriction of
$\psi$ on every connected component meets such conditions.

Let us take an additional divisor $Q=Q_1+\dots+Q_N$, on $\Gamma$, such that
$Q_i \in \Gamma \setminus \{P \cup D \cup R\}, i=1,\dots,N$ and let us put
$$
x^j(u^1,\dots,u^N)=\psi(u^1,\dots,u^N,Q_j), \ j =1,\dots,N.
$$
We have

\begin{theorem}[\cite{Krichever97}]
\label{krichever-theorem}
1) Let a holomorphic involution $\sigma:\Gamma\rightarrow\Gamma$
be defined on $\Gamma$ such that

a) $\sigma$ has exactly $2m$ fixed points, $m \leq N \leq 2m$:
$P_1,\dots, P_N$ and $2m-N$ points from $Q$;

b) $\sigma(Q)=Q$, i.e. for points from $Q$ the involution either interchanges them, either
let them fixed:
$$
\sigma(Q_k) = Q_{\sigma(k)}, \ \ \ k=1,\dots,N;
$$

c) $\sigma(k_i^{-1}) = -k_i^{-1}$ near $P_i$,
$i=1,\dots,N$;

d) there exists a meromorphic differential $\Omega$ on $\Gamma$
such that its divisors of zeros and poles are of the form
$$
(\Omega)_0= D + \sigma D +P, \ \ \ (\Omega)_{\infty}=R+\sigma R+Q;
$$

e) $\Gamma_0 = \Gamma/\sigma$ is a smooth algebraic curve.

\noindent
Then $\Omega$ is a pullback of some meromorphic differential
$\Omega_0$ on $\Gamma_0$ and we have the following equalities:
$$
\sum_{k,l} \eta_{kl}\frac{\partial x^k}{\partial u^i}
\frac{\partial x^l}{\partial u^j} =
\varepsilon^2_i H^2_i(u) \delta_{ij},
$$
where
$$
H_i = \lim_{P \to P_i} \left(\psi e^{-u^i k_i}\right), \ \ \
\eta_{kl} = \delta_{k,\sigma(l)} \res_{Q_k} \Omega_0,
$$
$$
\Omega_0 = \frac{1}{2} \left(\varepsilon_i^2 \lambda_i +
O(\lambda_i)\right) d\lambda_i, \ \lambda_i = k_i^{-2}, \
\mbox{at $P_i$, $i=1,\dots,N$}.
$$

2) If moreover there exists such an antiholomorphic involution
$\tau: \Gamma \to \Gamma$ such that all fixed points of $\sigma$
are fixed by $\tau$ and
$$
\tau^\ast(\Omega) = \overline{\Omega},
$$
then  $H_i(u)$ are real-valued for $u^1,\dots,u^N
\in \R$, and $u^1,\dots,u^N$ are orthogonal curvilinear coordinates
in the flat $N$-dimensional space with the metric $\eta_{kl}
dx^k dx^l$.
\footnote{It is easy to notice that if there are points
$Q_k$ and $Q_l, l =\sigma(k)$ which are interchanged by the involution then the metric
$\eta$ is indefinite.}

If all points from $Q$ are fixed by $\sigma$ and
\beq
\label{diff}
\res_{Q_1} \Omega_0 = \dots = \res_{Q_N} \Omega_0 = \eta^2_0 > 0,
\eeq
then the functions $x^1(u^1,\dots,u^N),\dots,x^N(u^1,\dots,u^N)$ solve the immersion problem
for orthogonal curvilinear coordinates $u^1,\dots,u^N$ and for the metric
$ds^2 = H_1^2 (du^1)^2 + \dots + H_N^2 (du^N)^2$ where
$$
H_i = \frac{\varepsilon_i h_i}{\eta_0}, \ \ \ i=1,\dots,N.
$$
\end{theorem}

{\sc Remark 1.}
Theorem still holds if instead of a) it is assumed that
the functions $\psi\exp(-f^i(u^i)k_i)$ are analytical near $P_i$ where $f^i$
are some functions of one variable, which are invertible near zero,
$i=1,\dots,N$. Therewith we do not differ orthogonal coordinates
which are obtained by change of variables of the form
$$
u^i\rightarrow f^i(u^i).
$$
It is also possible to change the condition
$\psi(u,R_k) = 1, k=1,\dots,l$,
by
\beq
\label{baker1}
 \psi(u,R_k) =d_k, \ \ k=1,\dots,l,
\eeq
where all constants $d_k$ do not vanish.
Moreover we even can assume only that
$d_k$ do not vanish simultaneously:
\beq
\label{baker2}
|d_1|^2 + \dots + |d_l|^2 \neq 0,
\eeq
and still Theorem holds.

For the case of orthognal coordinates in spaces of constant curvature Krichever's method
was recently generalized in \cite{BR}.

\subsection{Coordinate systems corresponding to singular spectral curves}
\label{subsec2.6}

Krichever's construction (Theorem \ref{krichever-theorem}) leads to theta-functional formulas
are difficult for qualitative analysis. At the same time after degenerating the spectral curve
$\Gamma$ we may obtain simpler formulas satisfying to
solutions which lie on the boundary of the moduli space found by Krichever.
Also for understand the structure of finite gap orthogonal curvilinear systems is
helpful to find among them classical coordinates.  The following theorem may serve for that.

\begin{theorem}[\cite{MT1}]
\label{th1}
1) Theorem \ref{krichever-theorem} holds for a singular algebraic curve
after replacing $g$ by $p_a(\Gamma)$, the arithmetic genus of $\Gamma$, and
replacing the smoothness condition d) for $\Gamma/\sigma$
by the condition that $P_1,\dots,P_N$ and the poles of $\Omega$ are nonsingular points.

Moreover we may assume that $\psi$ satisfies (\ref{baker1}) and (\ref{baker2}) instead of
the equalities $\psi(u,R_k)=1, k=1,\dots,l$.
\end{theorem}

In the case when $\Gamma_\nm$ is a union of smooth rational curves, i.e. of copies of
$\C P^1$, the constructions of the Baker--Akhiezer function and of orthogonal coordinates
reduce to simple computations with elementary functions and goes no further than solving
linear systems. We demonstrate that by examples.

We remark that the simplest spectral curves correspond to exotic coordinates
therewith for classical coordinates the spectral curves are rather complicated.

Let us recall that a regular differential $\Omega$ on $\Gamma$ is defined
by differentials $\Omega_1,\dots,\Omega_s$ on irreducible components
$\Gamma_1,\dots,\Gamma_s$. By its definition, we have
$\sum_i \sum_{\pi^{-1}(P) \in \Gamma_i} \ \Omega_i$ $=0$ where $\pi: \Gamma_\nm \to \Gamma$
is the normalization and the summation is taken over all singular points $P$.
The arithmetic genus $p_a$ is the dimension of the space of regular differentials.

{\sc Example 5.}
\footnote{We expose this simple example in details to
demonstrate the construction procedure, for details of other examples
we refer to \cite{MT1}.}
Let us consider the simplest singular spectral curve:
$\Gamma$ falls into two copies of $\C P^1$, i.e. $\Gamma_1$ and $\Gamma_2$,
which intersects at two points which implies $p_a(\Gamma)=1$: $a \sim b,
(-a) \sim (-b),  \{a,-a\} \subset \Gamma_1, \{b,-b\} \subset \Gamma_2$
(see Fig. 1). We consider the case with the essential singularities lying in
different irreducible components: at points
$P_1=\infty\in\Gamma_1$ and $P_2=\infty\in\Gamma_2$.
The Baker--Akhiezer function takes the form
$$
 \psi_1(u^1,u^2,z_1)=e^{u^1z_1}\left(f_0(u^1,u^2)+
\frac{f_1(u^1,u^2)}{z_1-\alpha_1}+\dots
 +\frac{f_k(u^1,u^2)}{z_1-\alpha_{s_1}}\right),\ z_1\in\Gamma_1,
$$
$$
 \psi_2(u^1,u^2,z_2)=e^{u^2z_2}\left(g_0(u^1,u^2)+
\frac{g_1(u^1,u^2)}{z_2-\beta_1}+\dots
 +\frac{g_n(u^1,u^2)}{z_2-\beta_{s_2}}\right),\ z_2\in\Gamma_2.
$$
$$
\psi_1(a)=\psi_2(b),\ \ \psi_1(-a)=\psi_2(-b).
$$

\vskip15mm

\begin{picture}(170,100)(-100,-80)
\qbezier(-10,0)(-10,30)(40,30) \qbezier(40,30)(90,30)(90,0)
\qbezier(90,0)(90,-30)(40,-30) \qbezier(40,-30)(-10,-30)(-10,0)
\put(35,33){\shortstack{$\Gamma_1$}}

\qbezier(60,0)(60,30)(110,30) \qbezier(110,30)(160,30)(160,0)
\qbezier(160,0)(160,-30)(110,-30) \qbezier(110,-30)(60,-30)(60,0)
\put(105,33){\shortstack{$\Gamma_2$}}

\put(-10,0){\circle*{3}} \put(90,0){\circle*{3}}
\put(60,0){\circle*{3}} \put(160,0){\circle*{3}}
\put(75,24){\circle*{3}} \put(75,-24){\circle*{3}}

\put(-25,0){\shortstack{$P_1$}} \put(165,0){\shortstack{$Q_2$}}
\put(95,0){\shortstack{$Q_1$}} \put(45,0){\shortstack{$P_2$}}
\put(63,20){\shortstack{\small{$a$}}}
\put(54,-26){\shortstack{\small{$-a$}}}
\put(85,18){\shortstack{\small{$b$}}}
\put(80,-26){\shortstack{\small{$-b$}}}
\put(60,-50){\shortstack{Fig. 1.}}
\end{picture}

The general normalization condition
is as follows
\beq
\label{normal}
\psi_1(R_{1,i})= d_{1,i}, \ \ \psi_2(R_{2,j})= d_{2,j},
\eeq
where
$R_{1,i}\in\Gamma_1, i=1,\dots,l_1$ и $R_{2,j} \in \Gamma_2, j=1,
\dots,l_2$. We also have
$l = l_1 + l_2 = s_1 + s_2$.
We assume that
$$
\Omega_1=\frac{(z_1^2-\alpha_1^2)\dots (z_1^2-\alpha_{l_1}^2)}
{z_1(z_1^2-a^2)(z_1^2-R_{1,1}^2)\dots(z_1^2-R_{1,l_1}^2)}\, dz_1,
$$
$$
\Omega_2=\frac{(z_2^2-\beta_1^2)\dots (z_2^2-\beta_{l_2}^2)}
{z_2(z_2^2-b^2)(z_2^2-R_{2,1}^2)\dots(z_2^2-R_{2,l_2}^2)}\, dz_2.
$$
Let us put $Q_1 = 0\in\Gamma_1, Q_2=0\in\Gamma_2$.
If the following inequalities
$$
\res_a \Omega_1 = - \res_b \Omega_2,\ \ \res_{-a}\Omega_1 =
-\res_{-b}\Omega_2,\ \ \res_{Q_1}\Omega_1 = \res_{Q_2}\Omega_2
$$
then the differential $\Omega$, defined by the differentials $\Omega_1$ and $\Omega_2$,
is regular and the condition (\ref{diff}) is satisfied. In this case, by
\ref{th1}, the coordinates $u^1$ and $u^2$ such that
$$
x^1(u) = \psi_1(u,0), \ \ x^2(u) = \psi_2(u,0),
$$
are orthogonal.

Let us consider the simplest case $l_1=0$ and $l_2 = 1$.
Then we have
$$
\psi_1=e^{u^1z_1}f_0(u^1,u^2),\ \
\psi_2=e^{u^2z_2}\left(g_0(u^1,u^2)+\frac{g_1(u^1,u^2)}{z_2-c}\right).
$$
The gluing conditions at the intersection points and the normalization condition
take the form $\psi_1(a)=\psi_2(b), \psi_1(-a)=\psi_2(-b), \psi_2(r)=1, r =
R \in \Gamma_2$ which imply
$$
\psi_1=e^{u^1z_1}
\frac{2b(c-r)e^{au^1+(b-r)u^2}}{(b+c)(b-r)e^{2bu^2}-(b+r)(b-c)e^{2au^1}},
$$
$$
\psi_2=e^{u^2z_2}\left(
\frac{e^{-ru^2}((b-c)e^{2au^1}+(b+c)e^{2bu^2})(c-r)}
{(b+c)(b-r)e^{2bu^2}-(b-c)(b+r)e^{2au^1}}+\right.
$$
$$
\left. \frac{1}{z_2-c}\frac{(b^2-c^2)(r-c)e^{-ru^2}(e^{2au^1}-e^{2bu^2})}
{(b+c)(r-b)e^{2bu^2}+(b-c)(b+r)e^{2au^1}}\right).
$$
On the components the differential $\Omega$ is defined by the differentials
$$
\Omega_1 = -\frac{dz_1}{z_1(z_1^2-a^2)}, \ \ \Omega_2 = -
\frac{(z_2^2-c^2)dz_2} {z_2(z_2^2-b^2)(z_2^2-r^2)},
$$
and their residues at singular points are equal to
$\res_{a}\Omega_1=\res_{-a}\Omega_1 = -\frac{1}{2a^2}= -
\res_{b}\Omega_2= -\res_{-b}\Omega_2=
\frac{(b^2-c^2)}{2b^2(b^2-r^2)}$.
The regularity condition (\ref{diff}) takes the form
$\res_{Q_1}\Omega_1= \frac{1}{a^2}= \res_{Q_2}\Omega_2 =
\frac{c^2}{r^2 b^2}$,
which implies
$a=\frac{br}{c},\ r=\frac{b}{\sqrt{2 + \frac{b^2}{c^2}}}$.
By straightforward computations we derive
$$
(x^1)^2+\left(x^2-e^{-ru^2}\frac{b(c-r)}{c(b^2-r^2)}\right)^2=
 e^{-2ru^2}\frac{b^2(c-r)^2}{c^2(b^2-r^2)^2}.
$$
The coordinate lines $u^2 = \const$ are circles centered at the
$x^2$ axis. For $b = \pm 1$ these circles touch the $x^1$ axis, and the second family of coordinate lines
$u^1 = \const$ consists in circles which are centered at the $x^1$ axis and which touch
the $x^2$ axis (see Fig. 2).

\vskip30mm

\begin{picture}(170,100)(-100,-80)
\put(60,-30){\vector(0,4){130}}
\put(-15,25){\vector(3,0){185}}

\put(60,35){\circle{20}} \put(60,45){\circle{40}}

\put(70,25){\circle{20}} \put(80,25){\circle{40}}

\put(160,30){\shortstack{$x^1$}} \put(65,92){\shortstack{$x^2$}}
\put(50,-50){\shortstack{Fig. 2.}}
\end{picture}

The case with the same spectral curve and the essential singularities lying in the
same component is considered in \cite{MT1}.

{\sc Example 6. The Euclidean coordinates.}
Let $\Gamma$ be a union of $n$ copies $\Gamma_1,\dots,\Gamma_n$ of the complex projective line
$\C P^1$. We put
$$
P_j=\infty, \ \ Q_j=0, \ \ R_j=-1\in\Gamma_j, \ \ \psi_j(R_j)=1, \ \
\ j=1,\dots,N.
$$
The differential $\Omega$ is given by the differentials
$\Omega_j=\frac{dz_j}{z_j (z_j^2-1)}$ on the components,  the Baker--Akhiezer function
$\psi$ equals $\psi_j=e^{u^jz_j}f_j(u^j)$, $j=1,\dots,N$,
and we obtain the Euclidean coordinates in $\R^N$: $x^j = e^{u^j}$.

{\sc Example 7. The polar coordinates.} The curve $\Gamma$ consists in five irreducible
components $\Gamma_1,\dots, \Gamma_5$, which do intersect as it is shown on Fig. 3.
We have $p_a(\Gamma)=1$.

Let us define an involution $\sigma$ on $\Gamma$ as follows:

а) the involution has the form $\sigma(z_j)=-z_j$,
on $\Gamma_1, \Gamma_2$, and $\Gamma_3$;

б) $\Gamma_3$ and $\Gamma_4$ are interchanged by $\sigma$ and
$b_1,c_1,\infty\in\Gamma_3$ are mapped into $b_2,c_2,\infty\in\Gamma_4$.

For a special choice of the divisors $D,P,Q$, and $R$ \cite{MT1}
these spectral data correspond to the polar coordinates:
$$
x^1 = \psi_5(Q_1) = r \cos \varphi, \ \ x^2 = \psi_5(Q_2) =  r \sin
\varphi, \ \ r = e^{u^1}, \ \ \varphi = u^2.
$$

\vskip30mm

\begin{picture}(85,50)(-150,-80)
\qbezier(-100,0)(-100,20)(-70,20) \qbezier(-70,20)(-40,20)(-40,0)
\qbezier(-40,0)(-40,-20)(-70,-20)
\qbezier(-70,-20)(-100,-20)(-100,0)

\qbezier(-50,0)(-50,40)(-20,40) \qbezier(-20,40)(10,40)(10,0)
\qbezier(10,0)(10,-40)(-20,-40) \qbezier(-20,-40)(-41,-37)(-46,-23)

\qbezier(7,39)(14,44)(30,45) \qbezier(30,45)(47,44)(54,39)
\qbezier(60,25)(60,5)(30,5) \qbezier(30,5)(0,5)(0,25)

\qbezier(0,-25)(0,-5)(30,-5) \qbezier(30,-5)(60,-5)(60,-25)
\qbezier(54,-39)(47,-44)(30,-45) \qbezier(30,-45)(14,-44)(7,-40)

\qbezier(50,0)(50,40)(80,40) \qbezier(80,40)(110,40)(110,0)
\qbezier(110,0)(110,-40)(80,-40) \qbezier(80,-40)(50,-40)(50,0)

\put(-100,0){\circle*{3}}

\put(-49,15){\circle*{3}}
\put(-43,15){\shortstack{\footnotesize$0$}}
\put(-57,7){\shortstack{\footnotesize$0$}}

\put(10,0){\circle*{3}} \put(14,-3){\shortstack{\footnotesize$P_2$}}

\put(50,0){\circle*{3}} \put(37,-3){\shortstack{\footnotesize$Q_1$}}

\put(110,0){\circle*{3}}
\put(113,-4){\shortstack{\footnotesize$Q_2$}}

\put(10,8){\circle*{3}} \put(2,3){\shortstack{\footnotesize$a$}}
\put(14,10){\shortstack{\footnotesize$b_1$}}

\put(10,-9){\circle*{3}} \put(-6,-9){\shortstack{\footnotesize$-a$}}
\put(14,-17){\shortstack{\footnotesize$b_2$}}

\put(50,8){\circle*{3}} \put(41,10){\shortstack{\footnotesize$c_1$}}
\put(54,1){\shortstack{\footnotesize$d$}}

\put(50,-9){\circle*{3}}
\put(41,-17){\shortstack{\footnotesize$c_2$}}
\put(53,-10){\shortstack{\footnotesize$-d$}}

\put(-113,0){\shortstack{\footnotesize$P_1$}}
\put(-72,25){\shortstack{$\Gamma_1$}}
\put(-25,45){\shortstack{$\Gamma_2$}}

\put(25,50){\shortstack{$\Gamma_3$}}
\put(25,-40){\shortstack{$\Gamma_4$}}
\put(78,45){\shortstack{$\Gamma_5$}}

\put(15,-60){\shortstack{Fig. 3}}

\end{picture}

{\sc Example 8. The cylindrical coordinates.}
The curve $\Gamma$ is a disjoint union of the curve $\widehat{\Gamma}$
from the previous example (the polar coordinates) and a copy $\Gamma_6$ of
$\C P^1$. All data related to $\widehat{\Gamma}$ are the same as before.
On $\Gamma_6$ we put $Q_3 = 0,
P_3 = \infty, R_4 = -1$
 и $\psi(R_4) = 1$.
Then we have $\psi_6(u^3)=e^{u^3(z_6+1)}$ and
$$
x^1 = \psi_5(Q_1) = r \cos \varphi, \ \ x^2 = \psi_5(Q_2) = r \sin
\varphi, \ \ x^3 = \psi_6(Q_3) = z,
$$
where $r = e^{u^1}, \varphi= u^2$ and $z = u^3$.

{\sc Example 9. The spherical coordinates in $\R^3$.}
The curve $\Gamma$ consists in $9$ irreducible components which intersect as it is shown on
Fig. 4. We have $p_a(\Gamma = 2$.

\vskip20mm

\begin{picture}(170,100)(-90,-80)

\qbezier(-85,0)(-85,20)(-55,20) \qbezier(-55,20)(-25,20)(-25,0)
\qbezier(-25,0)(-25,-20)(-55,-20) \qbezier(-55,-20)(-85,-20)(-85,0)

\put(-85,0){\circle*{3}} \put(-96,-4){\shortstack{\scriptsize$P_1$}}
\put(-65,25){\shortstack{\footnotesize$\Gamma_1$}}
\put(-42,18){\circle*{3}}
\put(-50,10){\shortstack{\footnotesize$0$}}
\put(-37,20){\shortstack{\footnotesize$0$}}

\qbezier(-45,0)(-45,40)(-15,40) \qbezier(-15,40)(15,40)(15,0)
\qbezier(15,0)(15,-40)(-15,-40) \qbezier(-15,-40)(-36,-37)(-41,-23)

\put(-15,45){\shortstack{\footnotesize$\Gamma_2$}}
\put(15,0){\circle*{3}} \put(15,7){\circle*{3}}
\put(15,-7){\circle*{3}} \put(18,-3){\shortstack{\scriptsize$P_2$}}
\put(7,2){\shortstack{\footnotesize$a$}}
\put(0,-7){\shortstack{\footnotesize$-a$}}
\put(18,8){\shortstack{\footnotesize$b_1$}}
\put(19,-15){\shortstack{\footnotesize$b_2$}}

\qbezier(7,39)(14,44)(30,45) \qbezier(30,45)(47,44)(54,39)
\qbezier(60,25)(60,5)(30,5) \qbezier(30,5)(0,5)(0,25)
\put(27,47){\shortstack{\footnotesize$\Gamma_3$}}
\put(45,7){\circle*{3}} \put(36,8){\shortstack{\footnotesize$c_1$}}

\qbezier(0,-25)(0,-5)(30,-5) \qbezier(30,-5)(60,-5)(60,-25)
\qbezier(54,-39)(47,-44)(30,-45) \qbezier(30,-45)(14,-44)(7,-40)
\put(27,-40){\shortstack{\footnotesize$\Gamma_4$}}
\put(45,-7){\circle*{3}}
\put(36,-15){\shortstack{\footnotesize$c_2$}}

\qbezier(45,0)(45,40)(75,40) \qbezier(75,40)(105,40)(105,0)
\qbezier(105,0)(103,-29)(97,-31) \qbezier(75,-40)(45,-40)(45,0)
\put(45,0){\circle*{3}} \put(33,-3){\shortstack{\scriptsize$Q_1$}}
\put(48,1){\shortstack{\footnotesize$d$}}
\put(47,-8){\shortstack{\footnotesize$-d$}}

\put(72,43){\shortstack{\footnotesize$\Gamma_5$}}
\put(90,36){\circle*{3}} \put(80,31){\shortstack{\footnotesize$0$}}
\put(97,31){\shortstack{\footnotesize$0$}}

\qbezier(75,0)(75,40)(105,40) \qbezier(105,40)(135,40)(135,0)
\qbezier(135,0)(135,-40)(105,-40) \qbezier(105,-40)(75,-40)(75,0)
\put(102,43){\shortstack{\footnotesize$\Gamma_6$}}
\put(135,0){\circle*{3}} \put(135,7){\circle*{3}}
\put(135,-7){\circle*{3}}
\put(139,-3){\shortstack{\scriptsize$P_3$}}
\put(126,2){\shortstack{\footnotesize$a$}}
\put(120,-7){\shortstack{\footnotesize$-a$}}
\put(138,9){\shortstack{\footnotesize$b_1$}}
\put(138,-15){\shortstack{\footnotesize$b_2$}}

\qbezier(127,39)(134,44)(150,45) \qbezier(150,45)(167,44)(174,39)
\qbezier(180,25)(180,5)(150,5) \qbezier(150,5)(120,5)(120,25)
\put(150,47){\shortstack{\footnotesize$\Gamma_7$}}
\put(165,7){\circle*{3}}
\put(155,8){\shortstack{\footnotesize$c_1$}}

\qbezier(120,-25)(120,-5)(150,-5) \qbezier(150,-5)(180,-5)(180,-25)
\qbezier(174,-39)(167,-44)(150,-45)
\qbezier(150,-45)(134,-44)(127,-40)
\put(150,-40){\shortstack{\footnotesize$\Gamma_8$}}
\put(165,-7){\circle*{3}}
\put(155,-14){\shortstack{\footnotesize$c_2$}}

\qbezier(165,0)(165,40)(195,40) \qbezier(195,40)(225,40)(225,0)
\qbezier(225,0)(225,-40)(195,-40) \qbezier(195,-40)(165,-40)(165,0)

\put(165,0){\circle*{3}} \put(152,-3){\shortstack{\scriptsize$Q_2$}}
\put(169,1){\shortstack{\footnotesize$d$}}
\put(168,-8){\shortstack{\footnotesize$-d$}}
\put(225,0){\circle*{3}} \put(228,-3){\shortstack{\scriptsize$Q_3$}}
\put(192,43){\shortstack{\footnotesize$\Gamma_9$}}

\put(75,-60){\shortstack{Fig. 4}}

\end{picture}

For certain choices of divisors $D,P,Q$ and $R$ (see \cite{MT1})
these spectral data lead to the spherical coordinates:
$$
x^1 = \psi_5(Q_1) = r \sin \varphi,\  \ x^2 = \psi_9(Q_2) = r \cos
\varphi \sin \theta,
$$
$$
x^3 = \psi_9(Q_3) = r \cos \varphi \cos
\theta, \ \ r = e^{u^1}, \ \varphi = u^2, \ \theta = u^3.
$$

{\sc Example 10. The spherical coordinates in $\R^N$.}
Let $\Gamma^{(N-1)}$ be the spectral curve and let $\psi^{(N-1)}$ the Baker--Akhiezer function
for $(N-1)$-dimensional spherical coordinates. The spectral curve
$\Gamma^{(N)}$ for the $N$-dimensional spherical coordinates is obtained from
$\Gamma^{(N-1)}$ and the curve from Fig. 5 via their intersection at $0\in\Gamma_{4N-7} \subset
\Gamma^{(N-1)}$ and $0\in\Gamma_{4N-6}$ (the number of irreducible components of
$\Gamma^{(k)}$ equals $4k-3$). We have $p_a(\Gamma^{(N)}) =
N-1$.

\vskip35mm

\begin{picture}(85,50)(-150,-80)

\qbezier(-50,0)(-50,40)(-20,40) \qbezier(-20,40)(10,40)(10,0)
\qbezier(10,0)(10,-40)(-20,-40) \qbezier(-20,-40)(-50,-40)(-50,0)
\put(-50,0){\circle*{3}} \put(-46,-3){\shortstack{\footnotesize$0$}}

\qbezier(7,39)(14,44)(30,45) \qbezier(30,45)(47,44)(54,39)
\qbezier(60,25)(60,5)(30,5) \qbezier(30,5)(0,5)(0,25)

\qbezier(0,-25)(0,-5)(30,-5) \qbezier(30,-5)(60,-5)(60,-25)
\qbezier(54,-39)(47,-44)(30,-45) \qbezier(30,-45)(14,-44)(7,-40)

\qbezier(50,0)(50,40)(80,40) \qbezier(80,40)(110,40)(110,0)
\qbezier(110,0)(110,-40)(80,-40) \qbezier(80,-40)(50,-40)(50,0)

\put(10,0){\circle*{3}} \put(14,-3){\shortstack{\scriptsize$P_n$}}

\put(50,0){\circle*{3}}
\put(28,-3){\shortstack{\scriptsize$Q_{n-1}$}}

\put(110,0){\circle*{3}} \put(113,-3){\shortstack{\scriptsize$Q_n$}}

\put(10,8){\circle*{3}} \put(3,2){\shortstack{\footnotesize$a$}}
\put(14,10){\shortstack{\footnotesize$b_1$}}

\put(10,-9){\circle*{3}} \put(-5,-8){\shortstack{\footnotesize$-a$}}
\put(14,-17){\shortstack{\footnotesize$b_2$}}

\put(50,8){\circle*{3}} \put(41,10){\shortstack{\footnotesize$c_1$}}
\put(54,1){\shortstack{\footnotesize$d$}}

\put(50,-9){\circle*{3}}
\put(41,-17){\shortstack{\footnotesize$c_2$}}
\put(53,-10){\shortstack{\footnotesize$-d$}}

\put(-27,45){\scriptsize{$\Gamma_{4n-6}$}}

\put(23,50){\scriptsize{$\Gamma_{4n-5}$}}
\put(23,-40){\scriptsize{$\Gamma_{4n-4}$}}
\put(76,45){\scriptsize{$\Gamma_{4n-3}$}}

\put(15,-60){\shortstack{Fig. 5}}

\end{picture}

It would be interesting to find the spectral data for other
known orthogonal coordinates and, in particular, for the elliptic coordinates
$u^1,u^2$ which are related to the Euclidean coordinates $(x^1,x^2)$
as follows
$$
x^1 = \cosh u^1 \, \cos u^2, \ \ \ x^2 = \sinh u^1\, \sin u^2.
$$

\subsection{A remark of discrete orthogonal coordinates}
\label{subsec2.7}

In an actively developing discretization of differential geometry by discrete coordinates in
$\R^N$ it is meant a mapping
$$
x: \Z^N \to \R^N
$$
which is an embedding of lattice. The translation operator $T_i$ along the $i$-th coordinate acts on
functions $F: \Z^N \to \R^k$ as follows
$$
T_i F(u^1,\dots,u^{i-1},u^i,u^{i+1},\dots,u^N) = F(u^1,\dots,u^{i-1},u^i+1,u^{i+1},\dots,u^N),
$$
and the partial derivation $\Delta_i$ in the $i$-th direction is defined as
$$
\Delta_i F(u) = T_i F(u) - F(u).
$$

By \cite{CDS}, a coordinate system is orthogonal if two conditions holds:

1) (the planarity condition) \ the points $\x(u), T_i\x(u), T_j\x(u), T_iT_j \x(u)$
lies in one plane for any given triple $i,j,u$;

2) (the circular condition) \ a planar polygon spanned by
$\x(u)$, $T_i\x(u)$, $T_j\x(u)$, and $T_i T_j \x(u)$ is inscribed into a circle.

In \cite{AVK} there was found a procedure for construction discrete Darboux--Egorov coordinates
(flat Egorov coordinates), based on the formal dis\-cre\-ti\-zation of the Baker--Akhiezer function
corresponding to continuous Darboux--Egorov coordinates from \cite{Krichever97}. Therewith
discrete Darboux--Egorov coordinates satisfy the planarity condition and the circular condition is
is strengthened as follows:

3) (the discrete Egorov condition) \ the edges $X^+_i(u) =
T_i\x(u)-\x(u)$ and $X^-_j(u) = T^{-1}_j\x(u) - \x(u)$ of the lattice are orthogonal
for all given triples $i,j,u$ (this implies that in any quadrangle from the circular condition two
right angles which are opposite to each other and therefore lean onto a diameter of the circle into which
this quadrangle is inscribed).

For discrete Darboux--Egorov coordinates there exists a discrete potential --- a function
$\Phi$ such that $\Delta_i \Phi(u) = |T_i\x(u) - \x(u)|^2$.

Analogously one may do a formal discretization (on the level of Baker--Akhiezer functions) of coordinates
given by Theorem \ref{th1}. Since everything is easily computed in this case, it is possible to find
explicitly examples of coordinates which, although do
satisfy the planarity condition, do not satisfy the circular condition.
This demonstrates the difficulties which appear in the study of an interesting problem consisting in finding a geometrically  meaningful discretization of the Lame equations.

\section{Frobenius manifolds}
\label{sec3}

\subsection{The associativity equations and Frobenius manifolds}
\label{subsec3.1}

Let us consider a finite-dimensional algebra generated by $e_1,\dots,e_n$
and with a commutative multiplication
$$
e_\alpha \cdot e_\beta = c^\gamma_{\alpha \beta} e_\gamma, \ \ \
c_{\alpha \beta \gamma} = \frac{\partial^3 F(t)}{\partial t^\alpha
\partial t^\beta \partial t^\gamma}, \ \ \ c^\gamma_{\alpha
\beta}=\eta^{\gamma\delta}c_{\alpha \beta \delta}, \ \
\eta^{\alpha\beta} = \eta^{\beta\alpha},
$$
where $F=F(t)$ is a function of $t=(t^1,\dots,t^N)$. The algebra is associative, i.e.
$$
(e_\alpha \cdot e_\beta)\cdot e_\gamma = e_\alpha \cdot (e_\beta
\cdot e_\gamma) \ \ \ \mbox{дл€ всех $\alpha,\beta,\gamma$,}
$$
if and only if $F$ satisfies {\it the associativity equations}
\beq
\label{e1}
\frac{\partial^3 F(t)}{\partial
t^\alpha
\partial t^\beta \partial t^\lambda} \, \eta^{\lambda \mu}
\frac{\partial^3 F(t)}{\partial t^\gamma
\partial t^\delta
\partial t^\mu} = \frac{\partial^3 F(t)}{\partial t^\gamma
\partial t^\beta \partial t^\lambda} \, \eta^{\lambda \mu}
\frac{\partial^3 F(t)}{\partial t^\alpha \partial t^\delta
\partial t^\mu}.
\eeq
In fact, if these conditions holds we obtain a family of associative algebras depending on an
$N$-dimensional parameter $t$.

These equations first appeared in the quantum field theory and
together with the following conditions:

1) $c_{1\alpha\beta} = \eta_{\alpha\beta}, \alpha,\beta=1,\dots,N$;

2) $\det (\eta^{\alpha\beta}) \neq 0$, $\eta^{\alpha\beta}\eta_{\beta\gamma} = \delta^\alpha_\gamma$ and
the metric $\eta_{\alpha\beta}$ is constant;

3) (the quasihomogeneity condition)
\beq \label{e2}
F(\lambda^{d_1}t^1,\dots,\lambda^{d_N}t^N) =
\lambda^{d_F}F(t^1,\dots,t^N)
\eeq

\noindent
they form the system of Witten--Dijkgraaf--Verlinde--Verlinde (WDVV) equations
\cite{W,DVV}.

There are two generalizations of the quasihomogeneity conditions:
it is assumed that there exists a vector field
$E = (q^\alpha
_\beta t^\beta + r^\alpha) \partial_\alpha$ which meets one of the following conditions:

3$^\prime$) the equality
$$
E^\alpha \partial_\alpha F = d_F F
$$
holds.
In (\ref{e2}) $E$ has the form $E = d_1 t^1 \partial_1 + \dots + d_N t^N
\partial_n$. This generalization covers the case of quantum cohomology;

3$^{\prime\prime}$) since by \cite{DVV} it is only important that the correlators
$c_{ijk}$, i.e. the third derivatives of  $F$, are homogeneous in the sense of
(\ref{e2}) it is sufficient to assume that
\beq
\label{quasi}
E^\alpha
\partial_\alpha F = d_F F + (\mbox{a second order polynomial in}\
t^1,\dots,t^N).
\eeq
This generalization is important for us because in our examples
some exponents $d_i$ are equal to $-1$.

A geometrical counterpart to solutions of the WDVV equations are Fro\-be\-nius
manifolds whose notion was
introduced by Dubrovin \cite{Dubrovin93}:
{\it a Fro\-be\-nius manifold} is a domain $U \subset \R^N$ endowed with a constant nondegenerate metric
$\eta_{\alpha\beta}du^\alpha du^\beta$ and with a solution to the associativity equations
({\it a prepotential}) $F$ satisfying the conditions 1), 2) and the most general quasihomogeneity condition
3$^{\prime\prime}$).

The quasihomogeneity condition came from physics and in addition to quantum filed theory Frobenius manifolds
play an important role in the theory of isomonodromic deformations
\cite{Dubrovin2008}.
\footnote{For isomonodromic deformations only
{\it semisimple manifolds}, i.e. without nilpotent elements, are interesting.}
However this condtions is the most restrictive and sometimes in the modern Frobenius geometry
it is omitted.

Before \cite{MT2} all know Frobenius manifolds were given by Dubrovin's examples of Frobenius structures
on the orbits of Coxeter groups (here the flat metric is the Sato metric; the
solutions the WDVV equations corresponding to singularities on type
$A_n$ were found in \cite{DVV}) and on the Hurwitz spaces, by quantum cohomology
and the extended moduli spaces of complex structures on Calabi--Yau manifolds
\cite{BK}, and by the ``doubles'' of the Hurwitz spaces found by Shramchenko
\cite{Sh}.

In every of these examples a Frobenius manifold has a special geometrical meaning. The examples from
\cite{MT2} are obtained by analytical methods (via finite gap integration), are algebraic in the sense that the correlators $c_{ijk} =
\frac{\partial^3 F}{\partial x^i
\partial x^j \partial x^k}$ are algebraic functions, and are not semisimple.

\subsection{Finite gap Frobenius manifolds}
\label{subsec3.2}

There is an important relation between solutions of the associativity equations and
flat Egorov metrics discovered by Dubrovin \cite{D0}. It is as follows. Let
$$
\eta_{\alpha\beta} dx^\alpha dx^\beta = \sum_{i=1}^N H^2_i(u)
\left(du^i\right)^2
$$
be a flat Egorov metric and $x^1,\dots,x^N$ be coordinates in some domain
which the coefficients $\eta_{\alpha\beta}$
are constant. We have
$$
\eta^{\alpha\beta} = \sum_{i=1}^N H^{-2}_i
\frac{\partial x^\alpha}{\partial u^i} \frac{\partial
x^\beta}{\partial u^i}
$$
and the conditions of symmetry for rotation coefficients and of zero curvature imply the existence of
a function $F(t)$ such that
 \beq
 \label{e4}
 c_{\alpha
\beta \gamma} = \sum_{i=1}^N H_i^2 \frac{\partial u^i}{\partial
x^\alpha} \frac{\partial u^i}{\partial x^\beta}\frac{\partial
u^i}{\partial x^\gamma} = \frac{\partial^3 F}{\partial x^\alpha
\partial x^\beta
\partial x^\gamma}
\eeq
and the associativity equations
$$
c^\lambda_{\alpha \beta} c^\mu_{\lambda\gamma} =
c^\mu_{\alpha\lambda} c^\lambda_{\beta\gamma} \ \ \ \mbox{for all}
\ \alpha,\beta,\gamma=1,\dots,N,
$$
holds where
$$
c^\alpha_{\beta\gamma} = \sum_i \frac{\partial x^\alpha}{\partial
u^i} \frac{\partial u^i}{\partial x^\beta}\frac{\partial
u^i}{\partial x^\gamma}.
$$
Assuming that the associative algebra is semisimple the converse is also true:
one may construct Egorov metric satisfying (\ref{e4}) from a solution
$F(t)$ of the associativity equations.

Since $u^1,\dots,u^N$ are orthogonal curvilinear coordinates in flat space,
one may apply Krichever's method for obtaining such coordinates
(see Theorem \ref{krichever-theorem}). In his article
\cite{Krichever97} there are given additional constraints on the spectral
data which correspond to Egorov metrics and therefore to solutions of the
associativity equations. However it sounds that from the properties of theta functions
one may derive that explicit theta-functional formulas obtained there do not
give quasihomogeneous solutions.

At the same time one may expect that solutions expressed in terms of elementary functions
and corresponding to singular curves may be quasihomogeneous and define
Frobenius manifolds. That was shown in
\cite{MT2}.

The following theorem distinguishes a special case when
the construction of Theorem \ref{th1} gives flat Egorov metrics and
quasihomogeneous solutions of the associativity equations.

\begin{theorem}[\cite{MT2}]
\label{th2}
1) Let the assumptions of Theorem \ref{th1} hold and
$\Gamma$ be a spectral curve such that all its irreducible components
are rational curves (complex projective lines $\C P^1$).

Let us pick on every component $\Gamma_i$ a pair of points $P_i = \infty,\ Q_i=0$  and
a global parameter $k_i^{-1} = z_i$,  $i=1,\dots,N$.
Let us assume that all singular points are double points of intersections of different components,
every such an intersection point
$a \in \Gamma_i \cap \Gamma_j$
has the same coordinates on both components:
$$
z_i(a)=z_j(a),
$$
and the involution $\sigma$ on every component has the form
$$
 \sigma(z_i)=-z_i.
$$
Then the metric
$$
ds^2 = \eta_{kl} d x^k d x^l = \sum_{i}
H^2_i \left(du^i\right)^2, \ \ H_i=H_i(u^1,\dots,u^N), \
i=1,\dots,N,
$$
constructed from this spectral data is a flat Egorov metric
(Darboux--Egorov metric).

2) Let us also assume that the spectral curve is connected and the Baker--Akhiezer function is normalized at one point $r$:
$$
\psi(u,r)=1, \ \ \ R = r \in \Gamma.
$$
Then the functions
$$
c_{\alpha \beta \gamma}(x)=\sum_{i=1}^n H_i^2\frac{\partial
u^i}{\partial x^\alpha} \frac{\partial u^i}{\partial
x^\beta}\frac{\partial u^i}{\partial x^\gamma},
$$
are homogeneous:
$$
c_{\alpha\beta\gamma}(\lambda x^1,\dots,\lambda x^n)=
\frac{1}{\lambda}c_{\alpha\beta\gamma}(x^1,\dots,x^n).
$$
\end{theorem}

To obtain from the derived solution of the associativity equations
a Frobenius manifold it needs to add to the generators $e_1,\dots,e_N$ of the associative algebra the unit
$e_0$ and a nilpotent element
$e_{N+1}$. Such an extension is given by the following algebraic lemma.

\begin{lemma}
[\cite{MT2}]
Let $F(t^1,\dots,t^N)$ be a solution of the associativity equations with
a constant metric $\eta_{\alpha\beta}$. Then the function
$$
\widetilde{F}(t^0,t^1,\dots,t^n,t^{N+1}) =
\frac{1}{2}\left(\eta_{\alpha\beta}t^\alpha t^\beta t^0 +
\left(t^0\right)^2 t^{N+1}\right) + F(t^1,\dots,t^N)
$$
satisfies the associativity equations (\ref{e1}) with the metric
$$
\widetilde{\eta} = \left(
\begin{array}{ccc}
0 & 0 & 1 \\
0 & \eta & 0 \\
1 & 0 & 0
\end{array}
\right) $$
and the associative algebra generated by
$e_0,e_1,\dots,e_N,e_{N+1}$ with the multiplication rules
$$
e_i \cdot e_j = c^k_{ij} e_k, \ \ \ c^k_{ij} = \widetilde{\eta}^{kl}
\frac{\partial^3 \widetilde{F}}{\partial t^l \partial t^i \partial
t^j},
$$
has the unit $e_0$:
$e_o \cdot e_k =e_k$ for all $k=0,\dots,n+1$,
and a nilpotent element $e_{N+1}$:
$e_{N+1}^2 = 0$.
Moreover if $F$ quasihomogeneous and $d_\alpha + d_\beta =c$ for all
$\alpha,\beta$ such that $\eta_{\alpha\beta} \neq 0$,  then
$\widetilde{F}$ is also quasihomogeneous with $d_0 = d_F-c, d_{N+1}= 2c -
d_F$ and the same values of $d_\alpha$, $\alpha=1,\dots,N$ as for
$F$.
\end{lemma}

Let us present two simplest examples of Frobenius manifolds
given by Theorem \ref{th2}.

We note that the statement 1) of this theorem gives solutions of the
associativity equations, and the statement 2) distinguishes a certain subclass of quasihomogeneous
solutions: there are quasihomogeneous solutions
which do not satisfy to the sufficient conditions form the statement 2).
Example 11 is covered by the statement 2), and Example 12 is not covered by it.

{\sc Example 11.} Let $\Gamma$ is formed by two complex projective lines $\Gamma_1$ and
$\Gamma_2$ which intersect at a pair of points (see Fig. 6):
$\{a,-a\in\Gamma_1\}\sim\{a,-a\in\Gamma_2\}$.

\vskip15mm

\begin{picture}(170,100)(-100,-80)
\qbezier(-10,0)(-10,30)(40,30) \qbezier(40,30)(90,30)(90,0)
\qbezier(90,0)(90,-30)(40,-30) \qbezier(40,-30)(-10,-30)(-10,0)
\put(35,33){\shortstack{$\Gamma_1$}}

\qbezier(60,0)(60,30)(110,30) \qbezier(110,30)(160,30)(160,0)
\qbezier(160,0)(160,-30)(110,-30) \qbezier(110,-30)(60,-30)(60,0)
\put(105,33){\shortstack{$\Gamma_2$}}

\put(-10,0){\circle*{3}} \put(90,0){\circle*{3}}
\put(60,0){\circle*{3}} \put(160,0){\circle*{3}}
\put(75,24){\circle*{3}} \put(75,-24){\circle*{3}}

\put(-25,0){\shortstack{$P_1$}} \put(165,0){\shortstack{$Q_2$}}
\put(95,0){\shortstack{$Q_1$}} \put(45,0){\shortstack{$P_2$}}
\put(63,20){\shortstack{\small{$a$}}}
\put(54,-26){\shortstack{\small{$-a$}}}
\put(85,18){\shortstack{\small{$$}}}
\put(80,-26){\shortstack{\small{$$}}} \put(60,-50){\shortstack{Fig. 6}}
\end{picture}

Let $N=2$, $l=1$ and the Baker--Akhiezer function
$\psi$ be normalized by the condition $\psi_2(r)=1$ at $r \in
\Gamma_2$. The prepotential takes the form
$$
F_{a,c}(x^1,x^2) = \frac{1}{4ac}\left(2x_2\sqrt{(a^2-c^2)x_1^2
+c^2x_2^2}\right.
$$
$$
+2cx_1^2\log \left(-\frac{cx_2+\sqrt{(a^2-c^2)x_1^2+c^2x_2^2}}
{x_1}\right)-\sqrt{2c^2-a^2}(x_1^2+x_2^2)
$$
$$
\left.\times\log\left(c^2(x_1^2-3x_2^2)+a^2(x_2^2-x_1^2)-
 2x_2\sqrt{2c^2-a^2}\sqrt{(a^2-c^2)x_1^2+c^2x_2^2}\right)\right)
$$
and satisfies the associativity equations for
$\eta_{\alpha\beta}=\delta_{\alpha\beta}$. It depends on two additional parameters
$a$ and $c$ and for $a=1, c=\frac{2}{\sqrt{7}}$ the formulas for the coordinates and for the
correlators are rather simple:
$$
x^1=\frac{4(7-\sqrt{7})e^{u^1-u^2}}{(21-6\sqrt{7})e^{2u^1}+(7+2\sqrt{7})e^{2u^2})},
$$
$$
x^2=\frac{e^{-2u^2}(3(\sqrt{7}-3)e^{2u^1}+(5+\sqrt{7})e^{2u^2})}
{3(\sqrt{7}-2)e^{2u^1}+(2+\sqrt{7})e^{2u^2}},
$$
$$
 c_{111}=-\frac{9x_1^8+51x_1^6x_2^2+88x_1^4x_2^4+(2x_1^2x_2^3+4x_2^5)\sqrt{(3x_1^2+4x_2^2)^3}+48x_1^2x_2^6}
 {2x_1(3x_1^4+7x_1^2x_2^2+4x_2^4)^2},
$$
$$
 c_{112}=\frac{9x_1^6x_2+15x_1^4x_2^3-8x_1^2x_2^5+(2x_1^2x_2^2+4x_2^4)\sqrt{(3x_1^2+4x_2^2)^3}-16x_2^7}
 {2(3x_1^4+7x_1^2x_2^2+4x_2^4)^2},
$$
$$
 c_{122}=-\frac{9x_1^7+15x_1^5x_2^2-8x_1^3x_2^4+(2x_1^3x_2+4x_1x_2^3)\sqrt{(3x_1^2+4x_2^2)^3}-16x_1x_2^6}
 {2(3x_1^4+7x_1^2x_2^2+4x_2^4)^2},
$$
$$
 c_{222}=\frac{-27x_1^6x_2-16x_2^7-72x_1^2x_2^5+(4x_1^2x_2^2+2x_1^4)\sqrt{(3x_1^2+4x_2^2)^3}-81x_1^4x_2^3}
 {2(3x_1^4+7x_1^2x_2^2+4x_2^4)^2}.
$$

{\sc Example 12.} Let $\Gamma$ be the same as in Example 11. In difference with Example 11
we assume that
$$
P_1 = \infty \in \Gamma_1, \ \ P_2 = 0\in \Gamma_1, \ \ Q_1 = \infty
\in \Gamma_2, \ \ Q_2 = 0 \in \Gamma_2,
$$
the normalization point $R=r$ lies in $\Gamma_1$ and the poles divisor $D=c$
lies in $\Gamma_2$ (see Fig. 7). However we do not assume that the intersection points have the same
coordinates (the assumptions of the statement 2) of Theorem \ref{th2}
are not satisfied):
$$
a \sim b, \ \ -a \sim -b, \ \ \ \pm a \in \Gamma_1, \ \pm b \in
\Gamma_2, \ a \neq b.
$$

\vskip15mm

\begin{picture}(170,100)(-100,-80)
\qbezier(-10,0)(-10,30)(40,30) \qbezier(40,30)(90,30)(90,0)
\qbezier(90,0)(90,-30)(40,-30) \qbezier(40,-30)(-10,-30)(-10,0)

\put(35,33){\shortstack{$\Gamma_1$}}

\qbezier(60,0)(60,30)(110,30) \qbezier(110,30)(160,30)(160,0)
\qbezier(160,0)(160,-30)(110,-30) \qbezier(110,-30)(60,-30)(60,0)

\put(105,33){\shortstack{$\Gamma_2$}}

\put(-10,0){\circle*{3}} \put(90,0){\circle*{3}}
\put(60,0){\circle*{3}} \put(160,0){\circle*{3}}
\put(75,24){\circle*{3}} \put(75,-24){\circle*{3}}

\put(-25,0){\shortstack{$P_1$}} \put(165,0){\shortstack{$Q_2$}}
\put(95,0){\shortstack{$P_2$}} \put(45,0){\shortstack{$Q_1$}}
\put(63,20){\shortstack{\small{$a$}}}
\put(54,-26){\shortstack{\small{$-a$}}}
\put(85,18){\shortstack{\small{$b$}}}
\put(80,-26){\shortstack{\small{$-b$}}}
\put(60,-50){\shortstack{Fig. 7}}

\end{picture}

The prepotential $F(x^1,x^2)$ equals
$$
F(x^1,x^2) =
-\frac{1}{8}\left(\left(x^1\right)^2+\left(x^2\right)^2\right)
\log\left(\left(x^1\right)^2+\left(x^2\right)^2\right)
$$
and is included in a linear family of
quasihomogeneous functions
$$
F_q(x^1,x^2) = q
\left(\left(x^1\right)^2+\left(x^2\right)^2\right)\arctan\left(\frac{x^1}{x^2}\right)
$$
$$
-\frac{1}{8}\left(\left(x^1\right)^2+\left(x^2\right)^2\right)
\log\left(\left(x^1\right)^2+\left(x^2\right)^2\right), \ \ q\in \R,
$$
which satisfy the associativity equations for
$\eta_{\alpha\beta}=\delta_{\alpha\beta}$.

The correlator for $F$ (i.e. $q=0$) has the simplest form:
$$
c_{111} =
-\frac{3}{2}\frac{x^1}{\left(x^1\right)^2+\left(x^2\right)^2} +
\frac{\left(x^1\right)^3}{\left(\left(x^1\right)^2+\left(x^2\right)^2\right)^2},
\ \
$$
$$
c_{112} =
-\frac{1}{2}\frac{x^2}{\left(x^1\right)^2+\left(x^2\right)^2} +
\frac{\left(x^1\right)^2 x^2
}{\left(\left(x^1\right)^2+\left(x^2\right)^2\right)^2},
$$
and the formulas for $c_{122}$ and $c_{222}$ are obtained by interchanging indices
$1 \leftrightarrow 2$.

Examples 11 and 12 describe quasihomogeneous deformations of the cohomology ring of
$\C P^2 \sharp \C P^2$. Indeed, there are the standard generators
$e_0,\dots,e_3$ in $H^\ast(\C P^2 \sharp \C P^2;\C)$:
$$
e_0 \in H^0, \ \ e_1, e_2 \in H^2, \ \ e_3 \in H^4, \ \
e_1^2 = e_2^2 = e_3, e_1e_2=0.
$$
We have $d_i = \frac{1}{2}\deg e_i$. The deformations change the multiplication
rules for two-dimensional classes by adding two-dimensional terms:
$$
e_i e_j = e_3 + c_{ij}^k(t) e_k, \ \ \ 1 \leq i,j \leq 2.
$$

\section{Soliton equations with self-consistent sources and the corresponding deformations of spectral curves}
\label{s4}

\subsection{Equations with self-consistent sources}
\label{subsec4.1}

In difference with $(1+1)$-dimensional soliton equations which have the Lax representation
$$
L_t = [L,A],
$$
equations with self-consistent sources are represented in the form
$$
L_t = [L,A] + C,
$$
where $C$ is an expression in terms of solutions $\psi_{k}$ of linear problems
$L\psi_k = \lambda_k\psi_k$, $k=1,\dots,l$.
To equations with self-consistent sources
there is applied the inverse problem method and it looks that
first, from the formal algebraical point of view, they were derived
by Melnikov \cite{Melnikov83}. In the physical literature first they did appear in the article
by Zakharov and Kuznetsov \cite{ZK} who, in particular, descried the physical meanings of
the KdV (see the formula (\ref{soliton2} below) and KP equations
$$
3u_{yy} - \frac{\partial}{\partial x}(u_t + 6uu_x +u_{xxx}) = 3\frac{\partial^2}{\partial x^2}|\psi|^2,
\ \ \
i\psi_t = \psi_{xx} + u\psi
$$
with self consistent sources.

Let us note two important facts:

1) every such an equation has as its predecessor a soliton equation
$L_t=[L,A]$;

2) for a correct posing the problem it needs to normalize the functions
$\psi$ (they have to satisfy to certain fixed spectral characteristics) and have the expression
$C$ which fits with the class of solutions in study --- for instance, to be fast decaying or periodic.

Solutions of such equations do have many interesting qualitative properties: for instance,
a creation and an annihilation of solitons do appear \cite{Melnikov89}.
We discuss that in в \S \ref{subsec4.3}.

In the periodic case to $L$ there corresponds a spectral curve which, as it appears, may be deformed
by an equation with self-consistent sources and therewith such a deformation consists in creation and annihilation
of double points. First this was observed in \cite{GT1} for the conformal flow which appears in differential geometry
of surfaces and later that was studied for the KP equation with self-consistent sources \cite{GT2}.
These two facts are exposed below in \S \ref{subsec4.2} and \S\ref{subsec4.3} respectively.

\subsection{Spectral curves of immersed tori and the conformal flow}
\label{subsec4.2}

In theory of the Weierstrass representation of immersed surfaces in
$\R^3$ and $\R^4$ (see the survey \cite{Taimanov2006}) surfaces are described in terms of solutions of the
equation $\D\psi=0$ for surfaces in $\R^3$ and of the equations $\D\psi = \D^\vee \varphi = 0$
for surfaces in $\R^4$ where
$$
\D =
\left(\begin{array}{cc} 0 & \partial \\
-\bar{\partial} & 0 \end{array} \right) +
\left(\begin{array}{cc} U & 0 \\
0 & \bar{U} \end{array} \right), \ \ \
\D^\vee =
\left(\begin{array}{cc} 0 & \partial \\
-\bar{\partial} & 0 \end{array} \right) +
\left(\begin{array}{cc} \bar{U} & 0 \\
0 & U \end{array} \right).
$$
For surfaces in $\R^3$ the potential $U$ is real-valued and
$\D = \D^\vee$.

For tori these operators are double-periodic and it is natural to consider their
spectral curves on the zero energy level. It appeared that these spectral curves contain
an information on the values of the Willmore functional
which is defined as follows:
$$
{\mathcal W} (M) = \int_{M} |{\bf H}|^2 d\mu,
$$
where ${\bf H}$ is the mean curvature vector and $d\mu$ is the volume form, on
$M$, induced by the immersion.
This functional is conformally invariant in the sense that
$$
{\mathcal W}(M) = {\mathcal W}(f(M)),
$$
where $M$ is a closed surface and $f: S^k \to S^k$ is a conformal transformation of
$S^k = \R^k \cup \{\infty\}$, $k\geq 3$, such that $M
\subset \R^k$ and $f(M) \subset \R^k$.

For tori of revolution the potential $U$ depends on one variable and
we obtain the reduction of the problem $\D \psi = 0$ to the Zakharov--Shabat spectral problem
for the one-dimensional Dirac operator.
Therewith the Willmore functional ${\mathcal W}$ becomes the first Kruskal--Miura integral for
the modified KdV (mKdV) hierarchy related to this operator.
Other Kruskal--Miura integrals of the mKdV  hierarchy are generalized to
(nonlocal) conservation laws for
the hierarchies related to the two-dimensional operator
$\D$ (the modified Novikov--Veselov equations and the Davy--Stewartson
equations).

This leads us to an observation on the existence of relation between
the geometry of surface and the spectral properties of the operator
$\D$ which comes into the Weierstrass representation of the surface
\cite{Taimanov1997}.

The approach to proving the Willmore conjecture which was proposed by us and which is based
on this relation until recently has not achieve a success.

Another our conjecture that higher two-dimensional generalizations
of the Kruskal--Miura integrals and together with them the whole
spectral curve are conformally invariant was confirmed in
\cite{GS}. For that there was used the conformal flow introduced by
Grinevich.

Since the conformal group of $\R^k \cup
\{\infty\}$ for $k \geq 3$ is generated by translations, dilations and
inversions, and since the potential
$U$ is invariant under translations and dilations, it is enough to prove
the invariance of spectral curve under inversions.
Let us consider the generator of inversions and
in the conformal group and let us write down its action of the potential.
It is as follows:
\beq
\label{conformal-flow-3}
U_t = |\psi_2|^2 - |\psi_1|^2,
\eeq
where a torus in $\R^3$ is defined by a solution
$\psi=(\psi_1,\psi_2)^\top$ of the equation $\D\psi=0$.

\begin{proposition}[\cite{GS}]
All Floquet multipliers of $\D$ are preserved by the flow
(\ref{conformal-flow-3}), which corresponds to an inversion of torus in $\R^3$,
there fore they are preserved by inversions.
\end{proposition}

It \cite{GT1} an analogous problem was considered for tori in $\R^4$ and it was showed that to the following
generator of inversions
$$
\partial_t x^1 = 2 x^1 x^3, \ \ \ \
\partial_t x^2 = 2 x^2 x^3,
$$
$$
\partial_t x^3 = (x^3)^2 -(x^1)^2 - (x^2)^2 - (x^4)^2,
\partial_t x^4 = 2 x^4 x^3
$$
there corresponds a deformation of $U$ of the form
\begin{equation}
\label{conformal-flow}
\partial_t U= \varphi_1\bar\psi_1-\bar\varphi_2\psi_2, \ \ \
\partial_t \bar U= \bar\varphi_1\psi_1-\varphi_2\bar\psi_2, \ \ \
\D \psi = \D^\vee \varphi = 0,
\end{equation}
where $\psi = (\psi_1,\psi_2)^\top$ and
$\varphi=(\varphi_1,\varphi_2)^\top$ define a torus in $\R^4$
via the Weierstrass representation.
Analogously to the three-dimensional case it was showed that

\begin{proposition}[\cite{GT1}]
All Floquet multiplies of $\D$ are preserved by the trans\-for\-mation, of
$U$, of the form (\ref{conformal-flow}) where
$\psi$ and $\varphi$ satisfy
$\D\psi = \D^\vee \varphi = 0$ and the quadratic expressions
$\psi_i^2,\varphi_j^2, i,j=1,2$, are double-periodic.
\end{proposition}

Let us note that the equations (\ref{conformal-flow-3})
and (\ref{conformal-flow}) are equations with
self-consistent sources which in the absence of these sources reduce to
the stationary equation $U_t=0$.

However there are known the explicitly computed spectral curves of the Clifford torus
$$
x_1^2 + x_2^2 = x_3^2 + x_4^2 = \frac{1}{2}
$$
in the three-sphere $S^3 \subset \R^4$
and of its stereographic projection into $\R^3$.
The Willmore conjecture stays that on the Clifford torus
the Willmore functional attains its minimal value
(for all immersed tori) which is equal to $2\pi^2$.

For the Clifford torus in $S^3 \subset \R^4$ the spectral curve is the complex projective line:
$\Gamma = \C P^1$,
and for its projection into $\R^3$ it is a rational curve with two double points \cite{Taimanov2005}.
A detailed analysis of the flow (\ref{conformal-flow}) led us to the following result.

\begin{theorem}[\cite{GT1}]
The evolution flow (\ref{conformal-flow}), acting on double-periodic po\-ten\-tials, preserve the Floquet
multipliers of $\D$ however it may deform its spectral curve (on the zero energy level)
and the deformation consists in creations and annihilations of double points.

Since the higher conservation laws for the modified Novikov--Veselov hie\-rar\-chy and for the
Davy--Stewartson hierarchy are expressed in terms of the asymptotics of multipliers,
the corresponding quantities are also preserved by the flow (\ref{conformal-flow}).
\end{theorem}

\subsection{Finite gap solutions of KdV and KP equations with self-consistent sources}
\label{subsec4.3}

Let $\Gamma$ be a Riemann surface of genus $g$ (the spectral curve) with a marked point
$P$ and with a local parameter $k^{-1}$, $k(P)=\infty$, near this point and let
there be defined a generic divisor $D = \gamma_1 + \dots + \gamma_g$, $\gamma_i \in \Gamma,
i=1,\dots,N$. Let us also assume that on $\Gamma$ there are marked $2N$ pair-wise different
$R_l^{\pm}, l =1,\dots,N$, which also differ from $P$ and from all points from $D$.

By the theory of Baker--Akhiezer functions, there exists a unique function
$\psi(\gamma,x,y,t,\tau)$,
$\tau=(\tau_1,\dots,\tau_N)$, $\gamma \in \Gamma$ such that

1) $\psi$ is meromorphic in $\gamma$ on $\Gamma \setminus P$ and has $g+N$ simple poles at
$\gamma_1,\dots,\gamma_g,R^+_1,\dots,R^+_N$;

2) $\res \psi(\lambda,x,y,t,\tau)\vert_{\lambda = R^+_l} = \tau_l \psi(R^-_l,x,y,t,\tau),\ l =1,\dots,N$,

3) $\psi$ has an essential singularity at and
$$
\psi(\gamma,x,y,t,\tau) = e^{k x + k^2 y + k^3 t} \left(1+ \sum_{m >0}
\frac{\xi_m(x,y,t,\tau)}{k^m}\right) \ \ \mbox{при $\gamma \to P$}.
$$

There exists the unique adjoint Baker--Akhiezer $1$-form
$\psi^\ast(\gamma,x,y,t,\tau)$
\footnote{The adjoint Baker--Akhiezer  $1$-forms were introduced in
\cite{Krichever82}.
The proof of its existence and uniqueness is analogous to proofs of the same facts of
Baker--Akhiezer functions}
with the following properties:

1$^\ast$) $\psi^\ast$ is meromorphic in $\gamma$ on $\Gamma \setminus P$ and has $g+N$ simple zeros at
$\gamma_1,\dots,\gamma_g,R^-_1,\dots,R^-_N$;

2$^\ast$) $\res \psi^\ast(\lambda,x,y,t,\tau)\vert_{\lambda = R^-_l} =
-\tau_l \frac{\psi^\ast(\lambda,x,y,t,\tau)}{d\lambda}\vert_{\lambda=R^+_l},\ l =1,\dots,N$,

3$^\ast$) $\psi^\ast$ has an essential singularity at $P$ and
$$
\psi^\ast(\gamma,x,y,t,\tau) = e^{-k x - k^2 y - k^3 t} (1+ o(1))\,dk \ \ \mbox{при $\gamma \to P$}.
$$

Let us put
\beq
\label{kpm}
u(x,y,t) = 2\frac{\partial}{\partial x} \xi_1(x,y,t,\tau), \ \ \ \tau_l = \alpha_l+\beta_l t,\ l=1,\dots,N.
\eeq
Analogously we construct from $\psi(\gamma,x,y,t,\tau)$ and $\psi^\ast(\gamma,x,y,t,\tau)$
the functions $\psi(\gamma,x,y,t)$ and the forms $\psi^\ast(\gamma,x,y,t)$.

We have

\begin{theorem}[\cite{GT2}]
The function $u(x,y,t)$ of the form (\ref{kpm})
satisfies the KP equation with $N$ self-consistent sources:
\beq
\label{kpmf}
u_t = KP[u] + 2\frac{\partial}{\partial x}\sum_{l=1}^N \beta_l \frac{\psi(R^-_l,x,y,t)\psi^\ast(\lambda,x,y,t)}{d\lambda}
\vert_{\lambda = R^+_l},
\eeq
where $u_t = KP[u]$ is the KP flow.

If there exists a two-sheeted covering $\pi:\Gamma \to \C P^1$, ramified at $P$, $\pi(P)=\infty$,
$\pi(k)=\pi(-k)$ and $\pi(R^+_l)=\pi(R^-_l)$, $l=1,\dots,N$,
then the formula (\ref{kpmf}) defines a solution $u(x,t)$ to the KDV equation with self-consistent sources
(in (\ref{kpmf}) it needs to replace $KP[u]$ by $KDV[u]$
 where $u_t=KDV[u]$ is the Korteweg--de Vries equation).
\end{theorem}

Analogously one may construct finite gap solutions of all equations of
the KdV and KP hierarchies with self-consistent sources \cite{GT2}.
The proof of this theorem is essentially based on the theory of
Cauchy--Baker--Akhiezer kernels introduced by Grinevich and Orlov \cite{GO}.

{\sc Remark 2.} The Baker--Akhiezer function $\psi$
from this theorem is defined on the spectral curve, with double points
$R_l$, which is obtained from $\Gamma$ by pair-wise identifications of $R^+_l$ and $R^-_l$,
$l=1,\dots,N$. The double point $R_l$ annihilates if and only if
$\tau_l=0$. If even for the initial potential $u(x,y,t)$, $t=0$,
the spectral curve is regular, then equations with self-consistent sources
immediately lead to creation of double points on the spectral curve
(for almost all times their number is equal to the number of sources $N$)
and these singularities are preserved for almost all times.

{\sc Example 13.} Let $\Gamma = \C \cup \{\infty\} = \C P^1$ with a parameter
$k$, $N=1$ and $R^\pm = \pm \kappa$. After simple computations we obtain the formula
$$
u(x,t,\tau) = -\frac{16
\tau\kappa^3}{(\tau e^{-(\kappa x + \kappa^3 t)} + 2\kappa e^{\kappa x
+ \kappa^3 t})^2},
$$
which gives a (regular) soliton for $\tau>0$,
a zero solution for $\tau=0$ and a singular soliton for $\tau <0$.
We derive that the function $u(x,t) = u(x,t,\alpha+\beta t)$
satisfies the KdV equation with a self-consistent source
\beq
\label{soliton2}
u_t = \frac{1}{4}\,u_{xxx} - \frac{3}{2}\,uu_x +
2\beta\partial_x
\psi^2(-\kappa,x,t),
\eeq
 where $\psi(-\kappa,x,t) = \left(1 - \frac{\tau}{\tau +2\kappa e^{2(\kappa x +\kappa^3 t)}}\right)e^{-\kappa x -\kappa^3 t}$, $ \tau = \alpha + \beta t$.
There the following qualitative effects first noticed by Melnikov
\cite{Melnikov89}:

\begin{enumerate}
\item
starting with a small initial value $c=c(0)$ we achieve $c=0$ in finite time
(an annihilation of soliton);

\item
by inverting the flow (\ref{soliton2}) for the initial data $c(0)=0$ we immediately
obtain a soliton for $t>0$ (a creation of soliton).
\end{enumerate}

For fast decaying potentials these effects are the analogs of
an annihilation and of a creation of double point for the spectral curves of double-periodic potentials.

\section{Some other examples of integrable problems with singular spectral curves}
\label{s5}

Let us expose a pair of interesting examples of integrable physical problems with singular spectral curves:

1) Recently Grinevich, Mironov and Novikov distinguished a class of algebro-geometrical spectral data from which by using two-point Baker--Akhiezer functions introduced in \cite{DKN} there are constructed operators of the form
$$
L = (\partial +A)\bar{\partial}
$$
(a magnetic Pauli operator) \cite{GMN1}.
The spectral curve $\Gamma$ reduces to two smooth components
$\Gamma^\prime$ and $\Gamma^{\prime\prime}$ from which is obtained by pair-wise gluing
$k+1$ pairs of points
$Q_1^\prime \sim Q_1^{\prime\prime}$, $\dots$, $Q_{k+1}^\prime \sim Q_{k+1}^{\prime\prime}$,
$Q^\prime_i \in \Gamma^\prime, Q^{\prime\prime}_j \in \Gamma^{\prime\prime}, i,j=1,\dots,k+1$.
Moreover there exists an antiholomorphic involution
$\sigma: \Gamma \to \Gamma, \sigma^2(\gamma)=\gamma, \gamma \in \Gamma$,
which interchanges the components:
$$
\sigma(\Gamma^\prime) = \Gamma^{\prime\prime}, \ \ \sigma(Q^\prime_k) = Q^{\prime\prime}_{\sigma(k)}.
$$
This implies that $\Gamma^\prime$ and $\Gamma^{\prime\prime}$ has the same genus $g$
and $p_a(\Gamma) = 2g+k$. Examples which are interesting from the physical point of view do appear
already in the case when
$\Gamma^\prime$ and $\Gamma^{\prime\prime}$ are complex projective lines..

With another real reduction of $L$:
$$
\widetilde{L} = (\partial_x + A)\partial_y,
$$
which is constructed from the same spectral curves,
there is related an integrable two-dimensional generalization of the Burgers equation \cite{GMN2}.

2) In the beginning of 1990s Krichever applied algebro-geo\-met\-ri\-cal me\-thods to a study of solutions of the
Yang--Baxter equations for  $(4 \times 4)$-matrices \cite{Krichever1981}. A construction of these solutions
does not use Baker--Akhiezer functions however the general ideology is taken from finite gap integration and the role of the spectral curve
$\Gamma$ is played by a smooth elliptic curve. Solutions are classified by their rank which takes values
$l=1,2$ and, as it is shown in \cite{Krichever1981}, all solutions of rank one are gauge equivalent
to the Baxter solution or obtained from it by transformations corresponding to simple symmetries.
Dragovich considered degenerated cases of this construction and showed that
if $\Gamma$ into two rational components which intersect at two points then we have
the well-known Yang solution \cite{D1} and if $\Gamma$ is a rational curve with a double point
then we get Cherednik's solution \cite{D2} (in both cases up to gauge equivalence).


\medskip
\addcontentsline{toc}{section}{References}

\begin{thebibliography}{MMM}

\bibitem{MT1}
Mironov, A.E., and Taimanov, I.A.:
Orthogonal curvilinear coordinate systems that correspond to singular spectral curves.
Proc. Steklov Inst. Math. 2006, no. 4 (255), 169--184.

\bibitem{MT2}
Mironov, A.E., and Taimanov, I.A.:
On some algebraic examples of Frobenius manifolds.
Theoret. and Math. Phys. {\bf 151} (2007), 604--613.

\bibitem{GT1}
Grinevich, P.G., and Taimanov, I.A.:
Infinitesimal Darboux trans\-formations of the spectral curves of tori in the four-space.
International Mathematics Research Notices 2007 (2007), rnm005, 1--21.

\bibitem{GT2}
Grinevich, P.G., and Taimanov, I.A.:
Spectral conservation laws for periodic nonlinear equations of the Melnikov type.
Amer. Math. Soc. Transl. Ser. 2, V. 224, 2008, P. 125--138.

\bibitem{Krichever97}
Krichever, I.M.:
Algebraic-geometric $n$-orthogonal curvilinear coordinate systems and the solution of associativity equations.
Funct. Anal. Appl. {\bf 31}:1 (1997), 25--39.

\bibitem{Novikov1974}
Novikov, S.P.:
A periodic problem for the Korteweg-de Vries equation.I.
Functional Anal. Appl. 8 (1974), no. 3, 236--246 (1975).

\bibitem{DKN}
Dubrovin, B.A., Krichever, I.M., and Novikov, S.P.:
The Schr\"odinger equation in a periodic field and Riemann surfaces.
Soviet Math. Dokl. {\bf 17} (1976), 947--952.

\bibitem{DMN}
Dubrovin, B.A., Matveev, V.B., and Novikov, S.P.:
Nonlinear equations of Korteweg--de Vries type, finite-band linear operators and Abelian varieties.
Russian Math. Surveys {\bf 31}:1 (1976), 59--146.

\bibitem{Krichever1977}
Krichever, I.M.:
Methods of algebraic geometry in the theory of nonlinear equations.
Russian Math. Surveys {\bf 32}:6 (1977), 185--213.

\bibitem{KN}
Krichever, I.M., and Novikov, S.P.:
Holomorphic bundles over algebraic curves, and nonlinear equations.
Russian Math. Surveys {\bf 35}:6 (1980), 53--80 (1981).

\bibitem{Dubrovin1981}
Dubrovin, B.A.:
Theta-functions and nonlinear equations.
Russian Math. Surveys {\bf 36}:2 (1981), 11--92 (1982).

\bibitem{Krichever1989}
Krichever, I.M.:
Spectral theory of two--dimensional periodic operators and its applications.
Russian Math. Surveys {\bf 44}:2 (1989), 145--225.

\bibitem{Taimanov2006}
Taimanov, I.A.:
The two-dimensional Dirac operator and the theory of surfaces.
Russian Math. Surveys {\bf 61}:1 (2006), 79--159.

\bibitem{Taimanov1998}
Taimanov, I.A.:
The Weierstrass representation of closed surfaces in $\R^3$.
Funct. Anal. Appl. {\bf 32}:4 (1998), 258--267 (1999).

\bibitem{Taimanov1997}
Taimanov, I.A.:
Modified Novikov--Veselov equation and differential
geometry of surfaces. Amer. Math. Soc. Transl., Ser. 2, V. 179,
1997, pp. 133--151.

\bibitem{Krichever1976}
Krichever, I.M.:
An algebraic-geometric construction of the Zakharov--Shabat equations and their periodic solutions.
Soviet Math. Dokl. {\bf 17} (1976), 394--397.

\bibitem{Serre}
Serre J.-P.:
Algebraic Groups and Class Fields, Graduate Texts in Math., {\bf 117}, Springer-Verlag, New York, 1988.

\bibitem{DKMM}
Dubrovin, B.A., Krichever, I.M., Malanyuk, T.M., and Makhankov, V.G.:
Exact solutions of the time-dependent Schro"dinger equation with self-consistent potentials.
Soviet J. Particles and Nuclei {\bf 19}:3 (1988), 252--269.

\bibitem{Taimanov2003}
Taimanov, I.A.:
On two-dimensional finite-gap potential Schr\"odinger and Dirac operators with singular spectral curves.
Siberian Math. J. {\bf 44} (2003), 686--694.

\bibitem{Malanyuk}
Malanyuk, T.M.:
A class of exact solutions of the Kadomtsev--Petviashvili equation.
Russian Math. Surveys {\bf 46}:3 (1991), 225--227.

\bibitem{DN83}
Dubrovin, B.A., and Novikov, S.P.:
Hamiltonian formalism of one-dimensional systems of the hydrodynamic type and
the Bogolyubov--Whitham averaging method.
Soviet Math. Dokl. {\bf 27} (1983), 665--669.

\bibitem{N85}
Novikov, S.P.:
Geometry of conservative systems of hydrodynamic type. The averaging method for field-theoretic systems.
Russian Math. Surveys {\bf 40}:4 (1985), 85--98.

\bibitem{DN89}
Dubrovin, B.A., and Novikov, S.P.:
Hydrodynamics of weakly deformed soliton lattices. Differential geometry and Hamiltonian theory.
Russian Math. Surveys {\bf 44}:6 (1989), 35--124.

\bibitem{Z}
Zakharov, V.E.:
Description of the $n$-orthogonal curvilinear
coordinate systems and Hamiltonian integrable systems of
hydrodynamic type, I: Integration of the Lam\'e equation. Duke
Math. J. {\bf 94} (1998), 103--139.

\bibitem{Tsarev}
Tsarev, S.P.:
Poisson brackets and one-dimensional Hamiltonian systems of hydrodynamic type. (Russian)
Soviet Math. Dokl. {\bf 31} (1985), 488--491.

\bibitem{Dubrovin90}
Dubrovin, B.A.:
On the differential geometry of strongly integrable systems of hydrodynamics type.
Funct. Anal. Appl. {\bf 24} (1991), 280-Ц285.

\bibitem{Pavlov}
Pavlov, M.V.:
Integrability of Egorov systems of hydrodynamic type.
Theoret. and Math. Phys. {\bf 150}:2 (2007), 225--243.

\bibitem{MF}
Mokhov, O.I., and Ferapontov, E.V.:
Nonlocal Hamiltonian operators of hydrodynamic type that are connected with metrics of constant curvature.
Russian Math. Surveys {\bf 45}:3 (1990), 218--219.

\bibitem{Darboux}
Darboux G.:
Lecons sur le System\'es Ortogonaux et Coordonn\'ees
Curvilignes, Gauthier--Villars, Paris, 1910.

\bibitem{ZM}
Zakharov, V.E., and Manakov, S.V.:
Reductions in systems integrable by the method of inverse scattering problem.
Doklady Matematics {\bf 57} (1998), 471--474.

\bibitem{BR}
Berdinsky, D.A., and Rybnikov, I.P.:
On orthogonal curvilinear coordinates in spaces of constant curvature.
Siberian Math. J. (to appear)

\bibitem{CDS}
Cieslinski, J., Doliwa, A., and Santini, P.M.:
The integrable discrete analogues of orthogonal coordinate systems are
multi-dimensional circular lattices, Phys. Lett. A. {\bf 235} (1997), 480--488.

\bibitem{AVK}
Akhmetshin, A.A.; Volvovskii, Yu.S., and Krichever, I.M.:
Discrete analogues of the Darboux--Egorov metrics.
Proc. Steklov Inst. Math. 1999, no. 2 (225), 16--39.

\bibitem{W}
Witten, E.:
On the structure of the topological phase of
two-dimensional gravity. Nucl. Phys. {\bf B 340} (1990), 281--332.

\bibitem{DVV}
Dijkgraaf, R., Verlinde, E., and Verlinde, H.:
Notes on topological
string theory and 2D gravity. Nucl. Phys. B {\bf 352} (1991),
59--86.

\bibitem{Dubrovin93}
Dubrovin, B.: Geometry of 2D topological field theories. Lecture
Notes in Math., {\bf 1620}, Springer, Berlin, 1995, 120--348.

\bibitem{Dubrovin2008}
Dubrovin, B.A.:
Hamiltonian partial differential equations and Frobenius manifolds.
Russian Math. Surveys {\bf 63}:6 (2008), 999--1010.

\bibitem{BK}
Barannikov, S., and Kontsevich, M.: Frobenius manifolds and
formality of Lie algebras of polyvector fields. Internat. Math. Res.
Notices (1998), no. 4, 201--215.

\bibitem{Sh}
Shramchenko, V.: "Real doubles" of Hurwitz Frobenius manifolds.
Comm. Math. Phys. {\bf 256} (2005), 635--680.

\bibitem{D0}
Dubrovin, B.: Integrable systems in topological field theory. Nucl.
Phys. B {\bf 379} (1992), 627--689.

\bibitem{Melnikov83}
Melnikov, V.K.:
Some new nonlinear evolution equations integrable by the inverse problem method.
Math. USSR-Sb. {\bf 49} (1984), 461--489.

\bibitem{ZK}
Zakharov, V.E., and Kuznetsov, E.A.:
Multi-scale expansions in the theory of systems integrable by the inverse
scattering transform, Physica D {\bf 18} (1986), 455--463.

\bibitem{Melnikov89}
Melnikov, V.K.:
Capture and confinement of solitons in nonlinear
integrable systems.  Comm. Math. Phys.  {\bf 120}  (1989),  no. 3, 451--468.

\bibitem{GS}
Grinevich, P.G., and Schmidt, M.U.:
Conformal invariant functionals
of immersions of tori into $\R^3$. J. Geom. Phys. {\bf 26} (1997),
51--78.

\bibitem{Taimanov2005}
Taimanov, I.A.:
Finite gap theory of the Clifford torus.
International Mathematics Research Notices (2005), 103--120.

\bibitem{Krichever82}
Krichever, I.M.:
The Peyerls model.
Functional Anal. Appl. {\bf 16}:4 (1982), 248--263 (1983).

\bibitem{GO}
Grinevich, P.G., and Orlov, A.Yu.:
Virasoro action on Riemann
surfaces, Grassmanians, $\det\bar\partial$ and Segal--Wilson
$\tau$-function. In: Problems of modern quantum field theory, Springer-Verlag, 1989,
86--106.

\bibitem{GMN1}
Grinevich, P.G., Mironov, A.E., and Novikov, S.P.:
Zero level of a purely magnetic two-dimensional nonrelativistic Pauli operator for
spin-$1/2$ particles.
Theoret. and Math. Phys. {\bf 164} (2010), 1110--1127.

\bibitem{GMN2}
Grinevich, P.G., Mironov, A.E., and Novikov, S.P.:
The two-dimensional Schr\"odinger operator: evolution $(2+1)$-systems and their new reductions,
the two--dimensional Burgers hierarchy, and inverse problem data.
Russian Math. Surveys {\bf 65}:3 (2010), 580--582.

\bibitem{Krichever1981}
Krichever, I.M.:
The Baxter equations and algebraic geometry.
Functional Anal. Appl. {\bf 15}:2 (1981), 92--103.

\bibitem{D1}
Dragovich, V.I.:
Solutions of the Yang equation with rational spectral curves.
St. Petersburg Math. J. {\bf 4} (1993), 921--931.

\bibitem{D2}
Dragovich, V.I.:
Solutions of the Yang equation with rational irreducible spectral curves.
Russian Acad. Sci. Izv. Math. {\bf 42}:1 (1994), 51--65.

\end{thebibliography}
\end{document}